\documentclass[fleqn,usenatbib]{mnras}

\usepackage{newtxtext,newtxmath}

\usepackage[T1]{fontenc}

\DeclareRobustCommand{\VAN}[3]{#2}
\let\VANthebibliography\thebibliography
\def\thebibliography{\DeclareRobustCommand{\VAN}[3]{##3}\VANthebibliography}

\usepackage{graphicx} 
\usepackage{amsmath} 
\usepackage{amssymb} 

\usepackage{longtable}
\usepackage{rotating}
\usepackage{lscape}
\usepackage{multicol}
\usepackage{multirow}
\usepackage{booktabs}
\usepackage{dirtytalk}
\usepackage{csquotes}

\title[Restoring the night sky at Observatorio del Teide]{Restoring the night sky darkness at Observatorio del Teide: First application of the model Illumina version 2}

\author[M. Aub\'e et al.]{Martin Aub\'e,$^{1,2,3}$ \thanks{E-mail: martin.aube@cegepsherbrooke.qc.ca (MA)}
Alexandre Simoneau,$^{3}$
Casiana Mu\~noz-Tu\~n\'on$^{4,5}$
\newauthor
Javier D\'iaz-Castro$^{4,5}$ 
and Miquel Serra-Ricart$^{4,5}$
\\
$^{1}$D\'epartement de physique, C\'egep de Sherbrooke, Sherbrooke,  475 rue du C\'egep, Sherbrooke, Qu\'ebec, J1E 4K1, Canada\\
$^{2}$Physics department, Bishops University, 2600 College St, Sherbrooke, Qu\'ebec, J1M 1Z7, Canada\\
$^{3}$D\'epartement de g\'eomatique appliqu\'ee, Universit\'e de Sherbrooke, 2500 Boulevard de l'Université, Sherbrooke, Qu\'ebec, J1K 2R1, Canada\\
$^{4}$Instituto de Astrof\'\i sica de Canarias, Calle Vía Láctea, s/n, 38205 San Cristóbal de La Laguna, Santa Cruz de Tenerife, 38205, Spain \\
$^{5}$Departamento de Astrof\'\i sica, Universidad de La Laguna (ULL), 38205 La Laguna, Tenerife, Spain}

\date{Accepted XXX. Received YYY; in original form ZZZ}

\pubyear{2020}

\begin{document}
\label{firstpage}
\pagerange{\pageref{firstpage}--\pageref{lastpage}}
\maketitle

\begin{abstract}
The propagation of artificial light into real environments is complex. To perform its numerical modelling with accuracy one must consider hyperspectral properties of the lighting devices and their geographic positions, the hyperspectral properties of the  ground reflectance, the size and distribution of small-scale obstacles, the blocking effect of topography, the lamps angular photometry and the atmospheric transfer function (aerosols and molecules). A detailed radiative transfer model can be used to evaluate how a particular change in the lighting infrastructure may affect the sky radiance.

In this paper, we use the new version (v2) of the Illumina model to evaluate a night sky restoration plan for the Teide Observatory located on the island of Tenerife, Spain. In the past decades, the sky darkness was severely degraded by growing light pollution on the Tenerife Island. In this work, we use the contribution maps giving the effect of each pixel of the territory to the artificial sky radiance. We exploit the hyperspectral capabilities of Illumina v2 and show how the contribution maps can be integrated over regions or municipalities according to the Johnson-Cousins photometric bands spectral sensitivities. 
The sky brightness reductions per municipality after a complete shutdown and a conversion to Light-Emitting Diodes are calculated in the Johnson-Cousins B, V, R bands. We found that the conversion of the lighting infrastructure of Tenerife with LED (1800K and 2700K), according to the conversion strategy in force, would result in a zenith V band sky brightness reduction of $\approx$ 0.3 mag arcsec\textsuperscript{-2}. 
\end{abstract}

\begin{keywords}
Light pollution -- Astronomical sites -- Radiative transfer model -- Night sky radiance - Lighting inventory
\end{keywords}



\section{INTRODUCTION}

The propagation of light in the nocturnal environment involves multiple physical interactions  \citep{aubephil2015}. In order to model that propagation with reasonable level of accuracy, one must include the information about the optical properties of the artificial light sources (spectral power distribution, angular emission) and their positions (latitude, longitude, elevation and height above ground). Other parameters such as the presence of blocking obstacles like trees and buildings, the spectral ground reflectance and the optical transfer function of the atmosphere (including aerosols) significantly influence light propagation  \citep{aube2007light,patat2008dancing,falchi2011campaign,pun2012night,pun2014contributions,puschnig2014night,aube2014evaluation,kyba2015worldwide,sanchezdemiguel2015variacion,Sanchez-sqmcolor}. 

The use of numerical models to study light pollution dates back to the 1980s with the Garstang model \citep{Garstang1986}. Numerical models allows a full control of the environmental parameters and provide the possibility to identify the origin of the light detected at a particular location in any viewing angle. The Garstang model contained many simplifying assumptions, mainly motivated by the low power of computers available at that time. Since then, many models have been developed with increased complexity to better describe the light pollution propagation in the nocturnal environment \citep{aube2005,aube2007light,kocifaj2007,luginbuhl2009,cinzano2013,baddiley2007,luginbuhl2009,aubephil2015,Falchie1600377,Aube2018}.

The Teide Observatory or Observatorio del Teide (OT) was founded in 1964. It is located in the Canary island of Tenerife at 2390 m of altitude. It is operated by the Instituto de Astrofísica de Canarias (IAC). It hosts many international telescopes and is the reference in solar astronomy. It benefits of good seeing conditions and good image quality \citep{vernin1992optical,vernin1994optical,munoz1997night,munoz2002astroclimatic,vernin2011european,perez2015forecasting}. Its artificial skyglow increased with the development of the touristic industry and with the general population increase. Its research capacities in the visible have been considerably reduced accordingly. For that reason, subsequent major optical telescopes were built at Roque de los Muchachos Observatory (ORM) on the nearby island of La Palma. In order to protect ORM sky from being altered too by light pollution, la \say{ley del cielo} \citep{law1988,law1992,law2017}, a national light pollution abatement, was voted by the Spanish government. This law comprises strict regulations to the lighting practices on the Island of La Palma but also some important restrictions for the Island of Tenerife. In 1997, the Spanish Government has subsidized the cost of a programme of street lighting replacement on La Palma island to minimize light pollution \citep{diaz1998}. The areas of Tenerife island facing ORM, hereafter called the protected area, experience more restrictive lighting rules than the rest of the island (hereafter called unprotected area). Thanks to that law, in the protected area, any lighting replacement has to be done using Phosphor Converted Amber light (PCamber) Light-Emitting Diodes (LED) with a reduction of output flux of 20\% (i.e., output flux is 0.8 of initial value). In the unprotected area, lighting replacement has to be done using 2700 Kelvin LEDs (LED2700K) with an output flux reduction of 70\% (i.e., output flux is 0.3 of its initial value). The smaller flux reduction for the protected area is explained by the fact that for this area, in the past, the allowed output flux was more restrictive than in the unprotected area. Basically, the protected area luminous flux was already reduced. At the end, when all light fixtures of the island will be converted to LEDs, both areas will have similar lighting levels. For the whole island of Tenerife, there is an additional flux reduction after midnight (output flux after midnight is 0.65 of the output flux before midnight).

The aim of this paper is to show up to what extent darkness of the sky around zenith at OT can be improved on the basis of its artificial sky radiance reduction. To achieve that, we first model the multispectral artificial sky radiance toward zenith and at 30 degrees from zenith for the  present situation. The crucial step to reach this first milestone was to define the lighting infrastructure and the obstacles properties all over the modelling domain. The modelled present artificial radiance is compared with All Sky Transmission Monitor (ASTMON, \citet{Aceituno2011}) Sky Brightness (SB) measurements in the B V and R Johnson-Cousins (JC) photometric bands for instruments installed at OT and ORM. Such a comparison is required in order to get a relevant estimate of the natural SB. This natural component comes from many sources like the starlight, the sky glow, the zodiacal light and so on. The natural SB and corresponding natural radiance are used to transform the calculation of the artificial sky radiance into the total SB (artificial + natural). In addition to modelling the present situation, two other modelled scenarios were performed to determine the effect of a full replacement of the light fixtures by:  1- PCamber, and 2- LED2700K. For these last two scenarios, we maintained the output flux equal to its present values. The results are weighted by the output flux reduction rules identified for the protected and unprotected areas.

\section{METHODOLOGY}

A simulation of the OT and ORM skies have been done by \citet{aube2012using} using version 0 of Illumina (v0). It was a comparison experiment with the MSNsRAu model \citep{kocifaj2007}. Among many differences with the version 2 (v2) used in the present paper, Illumina v0 was monochromatic and was using the Defense Meteorological Satellite Program - Operational Linescan System (DMSP-OLS, \citet{DMSP,elvidge1999radiance}) satellite data with much lower resolution and bad radiometric accuracy compared to Visible Infrared Imaging Radiometer Suite Day Night Band (VIIRS-DNB, \citet{elvidge2017viirs}) used in v2. It also had a crude correction for subgrid obstacles. 

There are not many methods to evaluate to what extent the sky quality of OT can be improved on the basis of its artificial sky radiance reduction. One of them can be that proposed by \citet{photonswithoutborders} where they compute the relative contribution of an area to the sky brightness of another. The present method is similar in the sense that we also integrate the contribution over municipalities, but since we are only concerned about a precise observing location we can model exactly the contributions to the sky radiance of the location using a complete radiative transfer model such as Illumina instead of relying on the use of uniform point-spread functions.
In this paper we are using the new version (v2) of the radiative transfer model Illumina to simulate the sky radiance in several wavelengths (spectral bins). Prior to the numerical calculations, it is important to define, as accurately as possible, the light fixture inventory, a list of the properties of the light sources (spectral power distribution and angular emission functions) and obstacles of the domain. This is probably the most difficult part of the work. As a result of the numerical calculations, we can exploit the modelled sky radiance in every spectral bin and combine them to create the artificial sky spectrum. We can also exploit the contribution maps. Such maps give the geographical distribution of the origin of the modelled sky radiance. There is one contribution map per spectral bin, per viewing angle and per lighting scenario. We integrated the contribution maps using the three JC bands (B, V, R) in order to compare them to the observed SB. The natural radiance in each JC band needs to be determined in order to calculate the total SB and radiance (natural + artificial). Contribution maps are also integrated over geographical limits of Tenerife municipalities and over the protected / unprotected area of the Tenerife Island. Such method is also applied to lighting conversion plans which allow the evaluation of the expected radiance reductions and SB decrease associated with the conversion or shutdown of each municipality or area.

\subsection{Illumina v2 model}

Illumina is a heterogeneous radiative transfer model dedicated to the simulation of the artificial sky radiance in any wavelength \citep{aube2005,aube2007light,aubephil2015,Aube2018}. The model is calculating the following physical interactions: 1- the aerosol (scattering and absorption) and molecular extinction (scattering only); 2- the 1\textsuperscript{st} and 2\textsuperscript{nd} order of scattering; 3- the ground reflection (lambertian); 4- the lamp flux; 5- the lamp angular emission function (horizontally averaged); 6- the topography; 7- the subgrid obstacles blocking (trees and buildings when the horizontal and vertical resolution cannot resolve them); 8- the reflection by overhead clouds. Illumina cannot yet calculate the molecular absorption. For that reason the use of Illumina must be restricted to the atmospheric windows but especially to the visible range. Since we used the newly released version of the model, it is worth highlighting the changes compared to the previous version (v1). The basic novelties of the model comprise

\begin{enumerate}
  \item An improvement of the calculation of the scattering probability and extinction. The probability of scattering is obtained from:  
     \begin{equation}
        p=1-\exp\left( \frac{\ln(T_\infty) \exp(-{z}/{H}) \mathrm{d}l }{H} \right)
     \label{prob}
     \end{equation}     
     Where $T_\infty$ is the vertical transmittance of the aerosols or molecules for the entire atmospheric  vertical column. $H$ is the scale height ($H=8$ km for molecules and 2 km for aerosols), $z$ is the altitude above ground and $\mathrm{d}l$ is the length of the scattering voxel. Similarly the transmittance of a light path is given by equations \ref{Thoriz} and \ref{Toblique} for a horizontal and an oblique light beam respectively:
     \begin{equation}
        T=\exp\left( \frac{\ln(T_\infty) \exp(-{z}/{H}) d }{H}  \right)
     \label{Thoriz}
     \end{equation}      
     \begin{equation}
        T=\exp\left( \frac{\ln(T_\infty)}{\cos(\theta_z)} \left[\mathrm{e}^{-z_a/H} - \mathrm{e}^{-z_b/H} \right] \right)
     \label{Toblique}
     \end{equation}
     Where $d$ is the horizontal distance of the light path, $z_a$ and $z_b$ are the bottom and top heights and $\theta_z$ is the zenith angle. $T_\infty$ for molecules ($T_{m\infty}$) is obtained using the extinction coefficient given by \citet{Kneizys1980} in their equation 18 (see equation \ref{Tm}  below). For aerosols, $T_{a\infty}$ is given by equation \ref{Ta}.
  \item The improvement of the accuracy in determining the solid angles. Especially when the scattering medium is located near the source, the first order scattering point, or the observer. In v1, the 3D space was divided into a fixed and coarse mesh grid, while in v2, we are defining small voxels on the fly. No vertical mesh grid is used anymore.
  \item The cloud base height can be set by the user, and a correction for the cloud fraction was added on the basis of \citet{scikezor2020impact} observations. We do not use this feature in this work since we are only concerned about clear skies.
  \item The addition of the direct radiance calculation. In v1, only the sky radiance was calculated. No direct sight to the light fixtures was allowed. This feature is not used in this work since we are focusing on the sky radiance. The direct radiance data are more suited for health and ecosystem studies.
  \item The Moderate Resolution Imaging Spectroradiometer (MODIS) reflectance product used in v1 is replaced by a weighted combination of surface reflectances to be defined by the user. This change has been implemented because 1) the low resolution of the MODIS data (500 m) that include many types of surface at the street-level scale so that the reflectance was not only representative of the ground below the light fixtures but rather of an average of surfaces, some lighted, some not; 2) the coarse resolution of VIIRS-DNB of 750 m do not allow the precise localization of the source and this is not enough accurate to identify the right reflectance to use even if we use high resolution reflectance data like the one from the Land Satellite (LANDSAT, \citet{masek2006landsat}); 3- satellite-based evaluation of the reflectance can be biased by obstacles that can hide the lighted surfaces and then introduce significant mismatch between the detected reflectance and the one of the surfaces underlying the lamp fixture. In v2 the reflectance is constant for all the modelling domain but has to be representative of the ground underlying the lighting devices. The ASTER spectral library \citep{baldridge2009aster} is routinely used for that purpose.
  \item The introduction of a multiscale grid that can allow a finer description of the environment near the observer. With this new feature, there is virtually no limit to the spatial resolution. In v1, the spatial resolution was fixed to 1 km. With v2, one can use very high resolution lidar data and then resolve the 3D buildings and trees effect on the light propagation. In v1, only a subgrid statistical obstacle correction was possible. Note that such statistical subgrid correction is still available in v2, depending on the resolution used in the multiscale definition of the modelling domain.
  \item The point source inventory can be directly used in the model to improve satellite-derived inventory. In v1, only satellite-derived inventory was possible. We do not use this feature in the present work.
\end{enumerate}

As for Illumina v1, Illumina v2 requires an \say{as accurate as possible} definition of a set of input data:

\begin{enumerate}
    \item angular emission function of the lamps\textsuperscript{*};
    \item spectra of the lamps\textsuperscript{*};
    \item lamp flux;
    \item lamp height relative to the ground;
    \item obstacles properties (height, distance, filling factor)\textsuperscript{*};
    \item underlying ground spectral reflectance;
    \item topography;
    \item minimum ground surface atmospheric pressure;
    \item relative humidity;
    \item $\tau_a$, Angstr\"om coefficient ($\alpha$) and the aerosol model
\end{enumerate}

Most of them are currently quite easy to define except the ones marked with an asterisk. Their determination requires the collaboration with a local expert that has good knowledge of the lighting infrastructure. We hope that with the rapid evolution of remote sensing techniques, having a local expert will not be required in a near future.

\subsection{Modelling experiments}

The aim of that work is to the evaluate the current level of light pollution and its possible change upon conversion of the lighting infrastructure with less polluting devices and better lighting practices. In that scope, it is very important to correctly define the geographical domain,  the lighting infrastructure, and environmental properties over that domain. In order to accurately model the contribution of the different municipalities of Tenerife Island on the sky radiance at OT we defined a finer resolution inventory for Tenerife while keeping a coarse definition for the other islands. It is well known, and since a long time, that the effect of light pollution is decreasing rapidly with distance \citep{bertiau1973artificial,treanor1973simple,berry1976light}. This fact stresses the importance of a better definition of the sources close to the observer. The experiment use 14 layers ranging from 20 m of resolution in the first central layer to a resolution of $\approx$ 1493 m for the 14\textsuperscript{th} layer. The resolution scale factor between two consecutive layers is 1.393. Given that, the resolution of the second layer is $\approx$28~m, the third $\approx$ 39~m and so on. The dimensions of each layer were 255x255 pixels. 

\begin{figure}
 \includegraphics[width=\columnwidth]{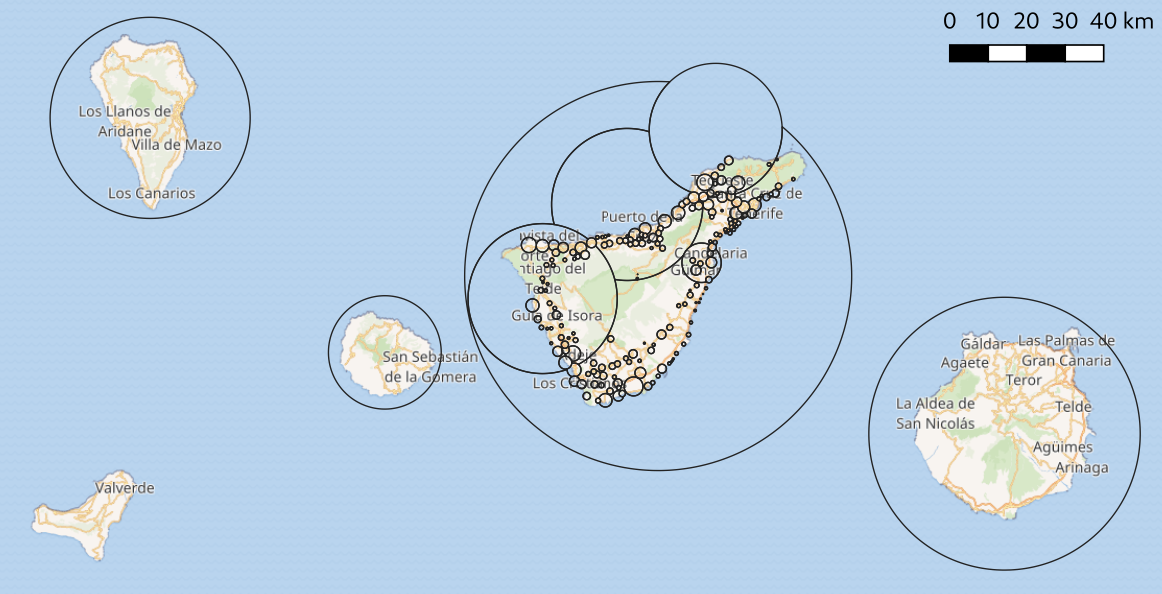}
    \caption{Circular zones used to define the properties of lighting devices and of the obstacles over the whole modelling domain. In the setting of the properties, the smaller zones overwrite the larger if ever there is an intersection between them. }
    \label{domain1}
\end{figure}

\begin{figure}
 \includegraphics[width=\columnwidth]{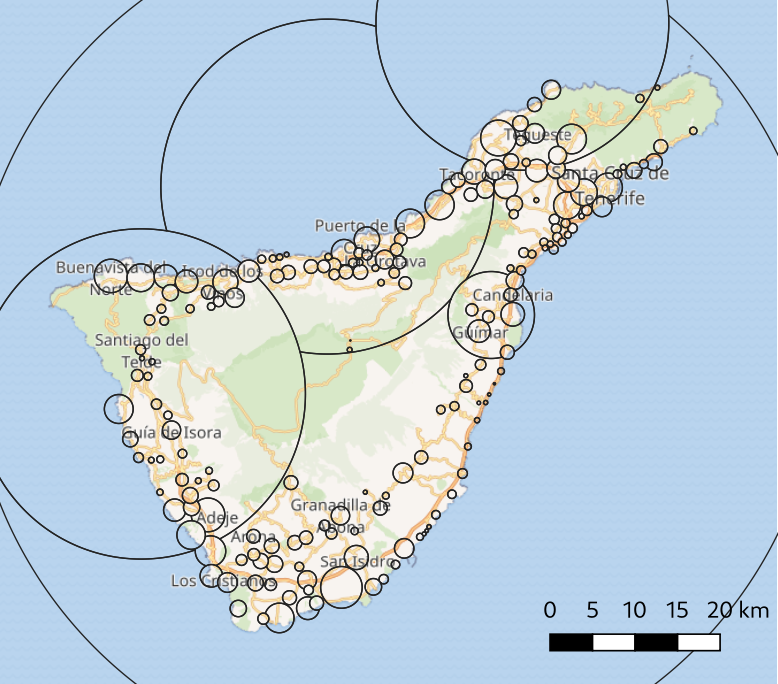}
    \caption{Enlargement of figure \ref{domain1} showing circular zones used to define the properties of lighting devices and of the obstacles over the Tenerife Island.}
    \label{domain2}
\end{figure}

Figures \ref{domain1} and \ref{domain2} show the various circular zones that were defined to characterize the different lighting and environment properties. On a given circular zone, we assume that the spectra, the angular emission functions, the lamp height and the obstacles properties are uniform. The light flux inside a given zone can vary given that it is derived using the VIIRS-DNB satellite monthly data (April 2019 in this study). Topography also varies inside a zone. The method used to convert VIIRS-DNB into flux is explained in \citet{Aube2018}. The complete set of data used for the inventory is given in Table \ref{inventory}.

\begin{figure}
 \includegraphics[width=\columnwidth]{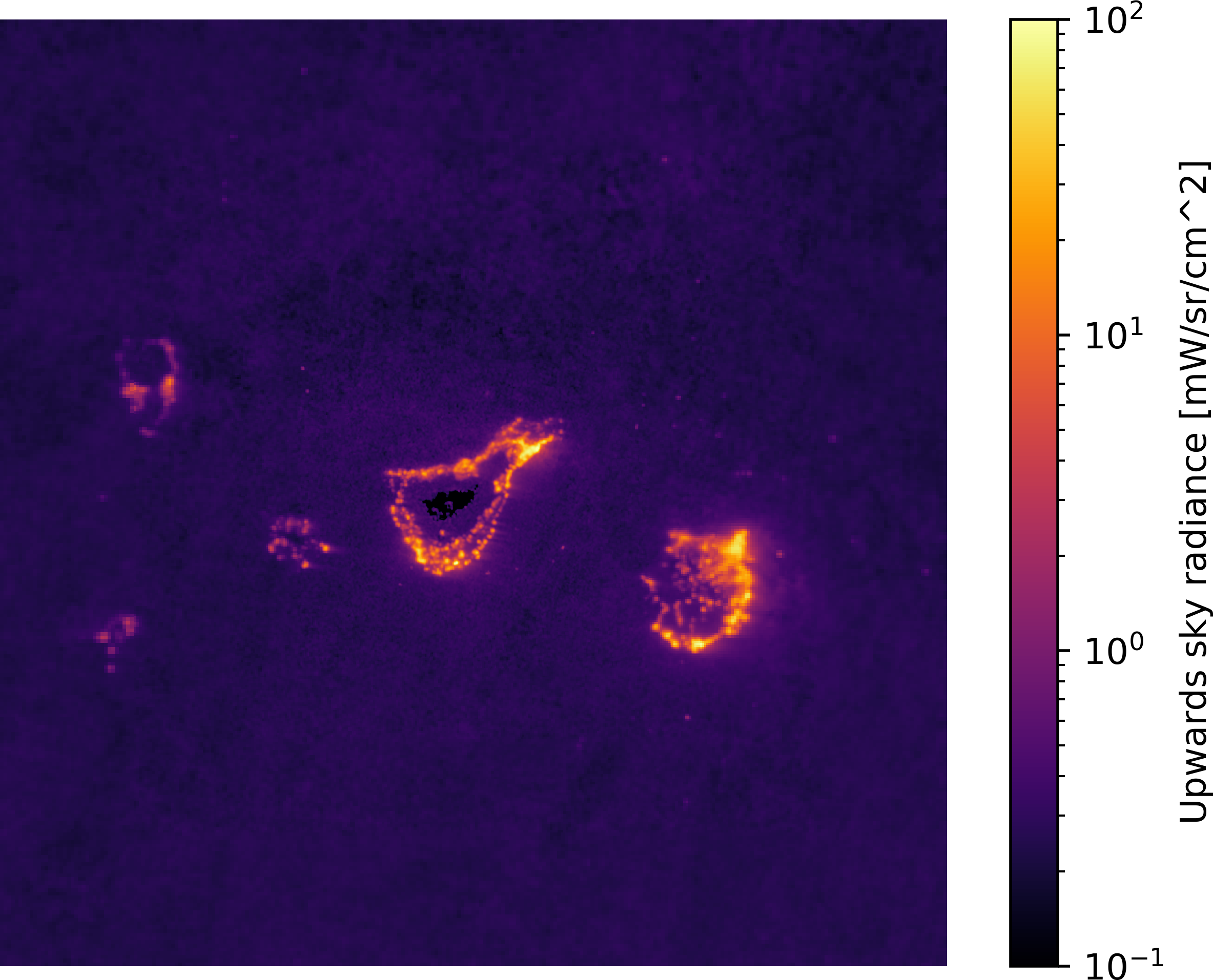}
    \caption{Original VIIRS-DNB radiances over the modelling domain.}
    \label{viirs_unfiltered}
\end{figure}

The modelling domain can be seen in Figure \ref{viirs_unfiltered}. This figure corresponds to the original VIIRS-DNB upward radiance data. The overall modelling domain covers $\approx$ 380 km E-W by $\approx$ 380 km N-S. The domain is centred on the observer position at OT (28.301197$^\circ$ N, 16.510761$^\circ$ W). We assume the observer to be 5 m above ground. 

\begin{figure}
 \includegraphics[width=\columnwidth]{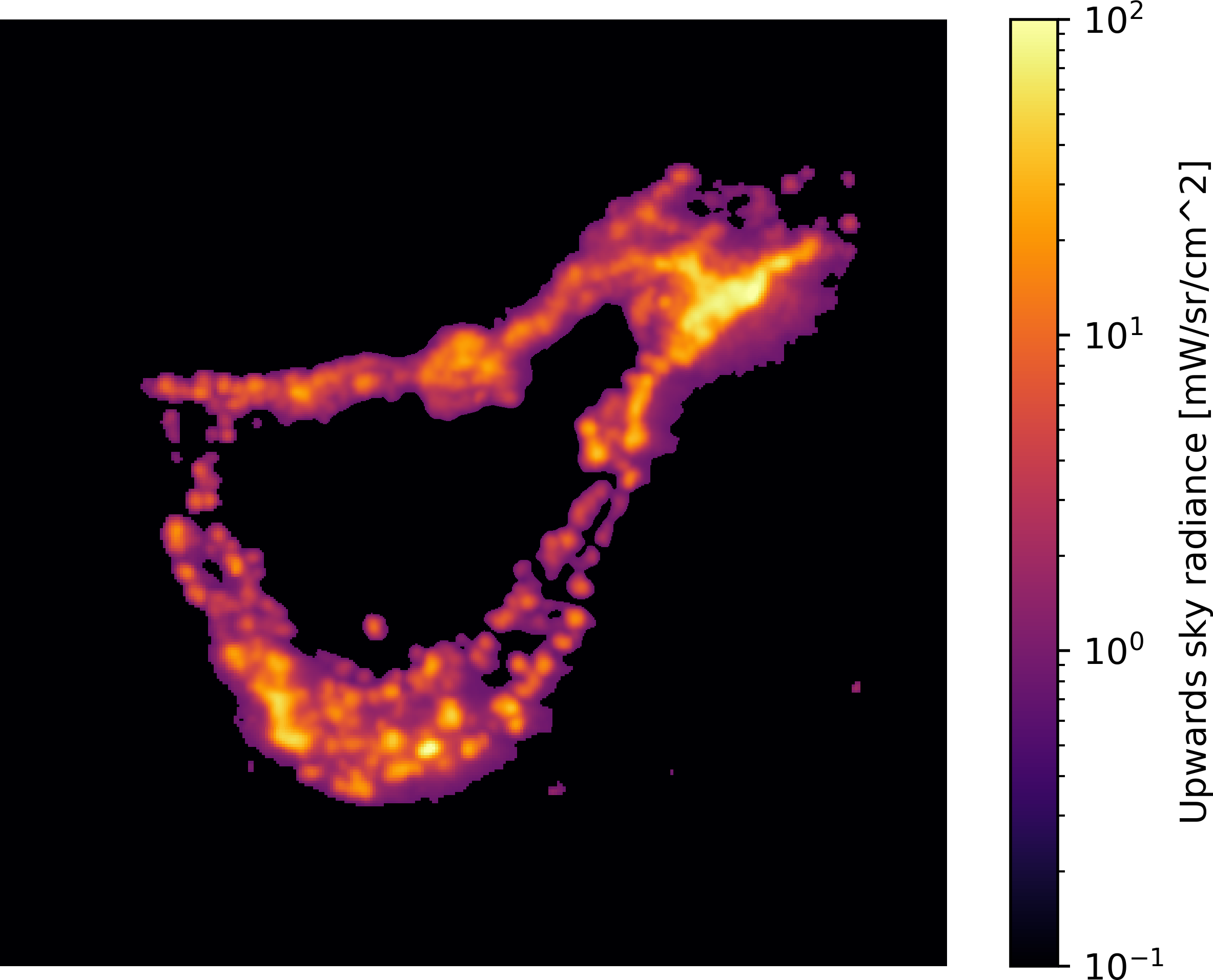}
    \caption{Filtered VIIRS-DNB data over the Tenerife Island.}
    \label{viirstenerife}
\end{figure}

A zoomed view on Tenerife Island is provided on Figure \ref{viirstenerife}. On that figure we filtered the original VIIRS-DNB radiances with a threshold of 0.8 nW/sr/cm$^2$. Such a threshold allowed to remove the background light over the ocean surface along with over the unlighted dense forest of the islands.

The calculations are made for 14 25 nm-wide spectral bins covering the spectral range of 380 nm to 730 nm. The sea-level air pressure is set to 101.3 kPa with an air relative humidity of 70\%. The sky is defined as cloudless. The maximum distance to calculate the effect of reflection on the ground is set to 9.99 m and the ground reflectance is defined by a weighted spectrum obtained assuming 90\% of asphalt and 10\% of grass. Reflectance spectra are taken from the ASTER spectral library \citep{baldridge2009aster}. We are using a $\tau_a$ at 500 nm of 0.04 and an angstrom coefficient of 1.1. Both values corresponding to the average of clear sky conditions for that period (April 2019) according to the Iza\~na sunphotometer of the Aerosol Robotic Network (AERONET, \citet{holben1998aeronet}). This Iza\~na sunphotometer is located only about one kilometer away from OT and is at almost the same altitude. We used the maritime aerosol model as defined by \citet{Shettle1979}.

We modelled three cases: 1- the present situation,  2- a complete conversion of the lamps to LED2700K, and 3- a complete conversion to PCamber. For both conversion scenarios, we first assumed that the output luminous flux was kept identical as it is in the present situation. At the end, to estimate the effect of real conversions, we weigh these results by their output luminous flux reductions according to the legal prescriptions described in the introduction (-20\% in the protected area and -70\% in the unprotected). In addition, we assume that replacement in the protected area is done using PCamber while LED2700K to be used in the unprotected area.

\subsection{Integration over geographical limits and JC bands}

\begin{figure}
 \includegraphics[width=\columnwidth]{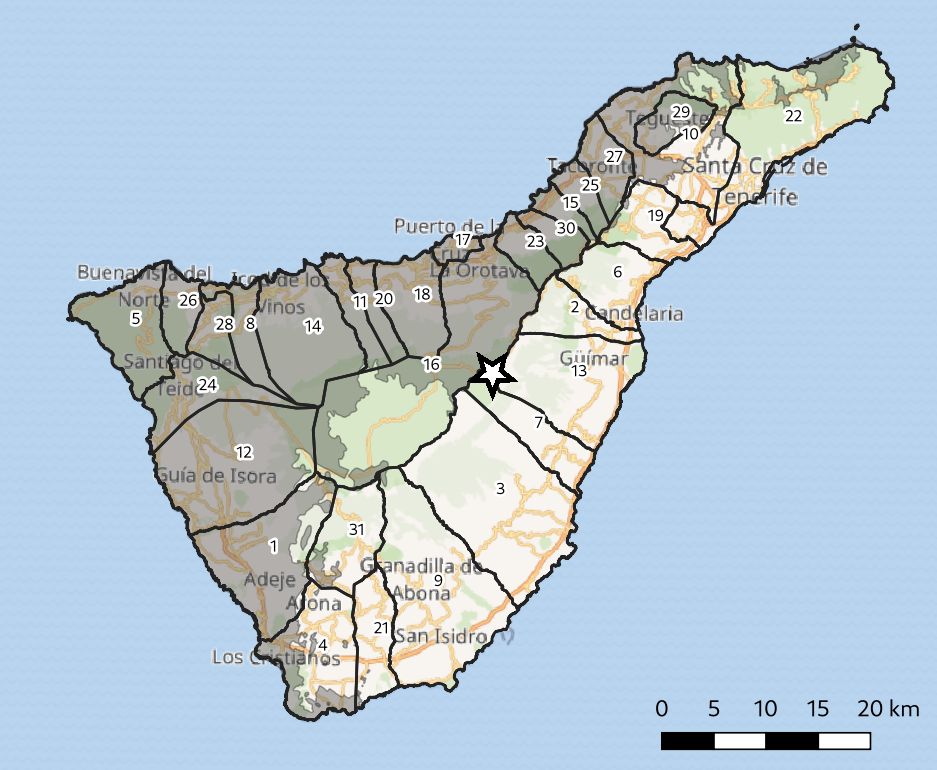}
    \caption{Limits of the various municipalities of Tenerife and protected (grey) / unprotected area. The observatory is marked by a star. Green color represent natural parks and reserves.}
    \label{municipios}
\end{figure}

There is one artificial radiance contribution map for each viewing angle and spectral bin. Since we are focused on the analysis of the sky brightness in the JC bands (B, V, R), we need to integrate the spectral information over each band. The process consists of doing the product of the spectral sensitivity of the JC band integrated over the spectral bin by the artificial radiance of that bin and integrate this quantity over the spectral range for each pixel of the modelling domain. The result is the JC band artificial radiance contribution map. Knowing the geographical limits of the municipalities and protected/unprotected areas, it is then possible to add up radiances of all pixel falling inside the limits of each municipality or area. Again we obtain an artificial radiance but it corresponds to the artificial radiance in JC bands for each municipality or area. The limits of the various municipalities and areas of Tenerife Island are illustrated in Figure \ref{municipios}. The star on that figure illustrates the position of the OT.

\subsection{Atmospheric and obstacle correction to the VIIRS-DNB inversion}
\label{corrections}

At the moment of writing this paper, Illumina do not correct for the VIIRS-DNB signal reduction caused by the atmospheric extinction and obstacles blocking. These corrections will soon be incorporated into the model. Their effect on each pixel of the modelling domain should be different because the obstacles and angular emission of light vary from one pixel to the other. In this study, we have made an approximate correction on the modelling output results instead of the modelling inputs. For the atmospheric correction, the correction is the same everywhere.  

To compensate for the molecular transmittance ($T_m$) and the aerosol transmittance ($T_a$), we use the following expression:
\begin{equation}
    F_T = \frac{1}{T_{a\infty} T_{m\infty}}
    \label{atm1}
\end{equation}

The molecular transmittance is calculated with equation \ref{Tm} derived from \citet{Kneizys1980}.
\begin{equation}
    T_{m\infty} = \exp \left( \frac{-1}{ \lambda^4 \left( 115.6406-\frac{1.335}{\lambda^2} \right)} \right)
    \label{Tm}
\end{equation}

$\lambda$ is in units of $\mu$m. The aerosol transmittance is calculated using
\begin{equation}
    T_{a\infty} = e^{-\tau_a}
    \label{Ta}
\end{equation}

Where $\tau_a$ is calculated at any wavelength ($\tau_a(\lambda)$) from $\tau_a$ at 500 nm ($\tau_a (0.5 \mu\mathrm{m})$) and with the Angstrom exponent $\alpha$.
\begin{equation}
    \tau_a(\lambda) = \tau_a (0.5 \mu\mathrm{m}) \left( \frac{\lambda}{0.5}  \right)^{-\alpha}
    \label{taua}
\end{equation}

$F_T$ can be easily estimated for the effective wavelength ($\lambda_e$) of the B, V, and R bands given in Table \ref{zeropoints}.

The buildings and trees are unresolved obstacles in the model but we include their statistical effects. However they are only considered to solve the radiative transfer but not to produce the input data. Obstacles blocking is determined by the average horizontal distance between the lamp and the obstacle ($d_o$), the average lamp height ($h_l$) and the average obstacle height ($h_o$) along with the obstacle filling factor ($f_o$). $f_o$ accounts for the fact that not all the light is intercepted by the obstacles, a part of it can pass through because there can be some space between the buildings and trees. These parameters were defined while building the inventory (see Table \ref{inventory}). In Table \ref{inventory}, \say{Obst. Distance} is equal to $2 \times d_o$. The VIIRS-DNB radiance monthly product is an average of radiances corresponding to a variety of zenith angles from 0 to $70^\circ$. But many of these angles are partly blocked by the subgrid obstacles. This blocking effect impacts in different ways the VIIRS-DNB radiance monthly product. There could be two components to the upward radiance: 1- the direct light and 2- the light reflected by the ground and obstacles surfaces. In most cases, the 2\textsuperscript{nd} component is the dominant one. This is because most light fixtures do not emit significantly at $\theta_z < 70^\circ$. If we define the obstacle correction factor as $F_{o}$, we can write the corrected radiance ${R_a}$ as a function of the uncorrected artificial radiance (${R_a}^*$) as follows:
\begin{equation}
    {R_a} \approx {R_a}^* F_T  F_{o}
    \label{retoile}
\end{equation}

If we assume that the street surface is the most lighted surface and then neglect the reflected light from the obstacles walls, we can define the limit zenith angle that allows reflected light to reach the satellite. This angle is given by:
\begin{equation}
    \theta_{lim} = \arctan \left(\frac{d_o}{h_o}\right)
    \label{anglim}
\end{equation}

The obstacles correction factor can be calculated by a weighting function of the solid angles.
\begin{equation}
    F_{o} \approx  \frac{\int_{0}^{70^\circ} \sin \theta_z \mathrm{d}\theta_z}{\int_{0}^{\theta_{lim}}  \sin \theta_z \mathrm{d}\theta_z + (1-f_o) \int_{\theta_{lim}}^{70^\circ}  \sin \theta_z \mathrm{d}\theta_z }
    \label{obsta0}
\end{equation}
\begin{equation}
    F_{o} \approx \frac{ 1 - \cos 70^\circ}{ 
    1- f_o \cos \theta_{lim} + (f_o-1) \cos 70^\circ }
    \label{obsta1}
\end{equation}

For typical Tenerife values of $h_o \approx$ 9 m, $d_o \approx 4$ m (i.e., about 8 m in diagonal between facing buildings), and $f_o \approx$ 0.9. This leads to $\theta_{lim} \approx $ 24$^\circ$ and then $F_o \approx 4.6$. 

As said, the obstacle correction varies from one pixel to another. For that reason the above correction is very approximate. We know that the obstacle correction is independent of the wavelength. It should be the same for the three Jonhson-Cousins bands. 

In this paper we decided to use sky brightness data acquired with ASTMON cameras during April 2019, both in OT and ORM, to determine if our $F_o$ estimate fit the observations and ultimately find a better value to use. We use the difference in the sky brightness in the V and B bands between the two sites. More specifically we use the 75 percentile (P75) values (see table \ref{sbmeas}). P75 values  were estimated to provide the best proxy of the darkest conditions during April 2019. Using 99 percentile (P99) data should normally be better but, for the month of April 2019, there was some contamination in the P99 band data that disappeared in the 75 percentile data. We assume that the sky brightness at ORM in its best atmospheric conditions, is very close to the natural sky brightness. This assumption should be true within 0.03 mag arcsec\textsuperscript{-2} according to \citet{benn1998palma}. This assumption do not apply to the R band.  We exclude the R band because that, on La Palma Island, many of the light fixtures are either Low-pressure sodium or monochromatic amber LEDs. These artificial lights only emit in the R band. For that reason, we cannot assume that the R band sky brightness at ORM is as representative of the natural sky brightness as the B and V bands. Table \ref{sbmeas2018-2019} show the P99 zenith sky brightness recorded at ORM for the years 2018-2019. These measurements are brighter than the natural sky brightness $S_{bg}$ estimates made by \citet{benn1998palma} in the B and V (-0.07 mag arcsec\textsuperscript{-2} in B, -0.09 mag arcsec\textsuperscript{-2} in V). But we recall that \citet{benn1998palma} suggested to add 0.03 mag arcsec\textsuperscript{-2} in all bands to determine the natural level. This is what we have done in B and V. Considering that, the new 2018-2019 measurements are consistent with the 1998 measurements in B and V bands. In the R band, the P99 measurement is darker than the \citet{benn1998palma} estimate (+0.15 mag arcsec\textsuperscript{-2}). The ORM sky brightness decrease in the R band since 1998 is probably due to the significant change in the lighting systems on La Palma. In 1998, there was a lot of Low Pressure Sodium lamps that are emitting in the R band. Then we must admit that the natural SB evaluation made by \citet{benn1998palma} in the R band was overestimated of at least 0.15 mag arcsec\textsuperscript{-2}. For that reason we will use the 2018-2019 P99 data as the best estimate of $S_{bg}$ in the R band while keeping the \citet{benn1998palma} values for B and V bands.

\begin{table}
 \centering
 \caption{Percentile 75 ASTMON sky brightness measurements and differences in April 2019 at OT and ORM. The $\Delta_S$ values of  that table are used to determine precisely the correction factor $F_o$.}
 \label{sbmeas}
 \begin{tabular}{ccccc }
 \hline
Band & Product & S (OT) & S (ORM) & $\Delta_S$ \\
 & & mag arcsec\textsuperscript{-2} & mag arcsec\textsuperscript{-2} & mag arcsec\textsuperscript{-2} \\
\hline
B & P75 & 22.11 & 22.26 & 0.15 \\
V & P75 & 21.06 & 21.52 & 0.46 \\
R & P75 & 20.79 & 20.83 & 0.04 \\
  \hline
 \end{tabular}
\end{table}

\begin{table}
 \centering
 \caption{Percentile 99 ASTMON sky brightness measurements for 2018-2019 at ORM.}
 \label{sbmeas2018-2019}
 \begin{tabular}{ccccc }
 \hline
Band	&	Product	&	$S$ (ORM)	&	Error	&	nb data	\\
	&		&	mag arcsec\textsuperscript{-2}	&	mag arcsec\textsuperscript{-2}	&	-	\\
	\hline
B	&	P99	&	22.66	&	0.03	&	814	\\
V	&	P99	&	21.84	&	0.02	&	762	\\
R	&	P99	&	21.18	&	0.01	&	1,157	\\
  \hline
 \end{tabular}
\end{table}

The obstacles correction factor was empirically verified so that the total  modelled zenith sky brightness reduction upon a complete shutdown of the lights (see Table \ref{magshutdown}) fit with the measured SB differences in B and V bands between OT and ORM (see Table \ref{sbmeas}). This exercise led us to a value of $F_o=$ 5.05 instead of 4.6 (i.e., 10\% larger). With that empirical value, we obtain a fit of B and V bands reductions that is within 0.02 mag arcsec\textsuperscript{-2}. Part of this correction can come from the fact that VIIRS-DNB data are not acquired at the same moment than SB measurements. But it is most probably coming from the use of only one set of obstacles values to estimate $F_o$ that may not correctly represent the influence of all the pixels of the domain to the modelled sky brightness. This is why we should implement this  directly to the input data in the future. The obstacle model used only assumes a single layer of uniform obstacles. But we assume that a more complete description such as the one made by \citet{Kocifaj2018} could not provide a significant improvement to the correction of the VIIRS-DNB signal because of the relatively low zenith angles considered ($\theta_z \leq 70^\circ$).

\begin{table}
 \centering
 \caption{Correction factors to the modelled radiance assuming $h_o$=9, $f_o$=0.9, and $d_o$=6.}
 \label{corrfactor}
 \begin{tabular}{cccccc } 
  \hline
  Band & $\lambda_{e}$ & $T_m$ & $T_a$ & $F_T$ & $F_{o}$  \\
                          & nm & - & - & - \\
     \hline                      
        B &  436.1 & 0.775 & 0.922 & 1.40 & 4.6  \\
        V &  544.8 & 0.903 & 0.938 & 1.18 & 4.6  \\
        R &  640.7 & 0.949 & 0.948 & 1.11 & 4.6  \\
  \hline
 \end{tabular}
\end{table}

\subsection{Conversion from radiance to sky brightness}

Illumina calculates only the artificial sky radiance. In order to determine the total SB equivalent in units of mag arcsec\textsuperscript{-2} it is mandatory to get a good estimate of the natural component of the SB for the site and period.  The natural SB is highly variable with time, altitude, season and observing direction. It is composed of light from multiple sources such as the zodiacal light, the starlight, the sky glow and the Milky Way \citep{benn1998palma}. In this study we are using natural sky brightness estimates made by \citet{benn1998palma} to determine the background SB and radiance. This natural sky brightness excludes starlight. For that reason, the ASTMON measurements are shifted compared to the background SB, but the shift is the same for both OT and ORM when considering the same period. This is why we used the SB differences between the two sites instead of absolute values to calibrate the model results as explained in Section \ref{corrections}.

For a given JC band, let’s call the radiance responsible for the natural contribution without starlight, the background radiance ($R_{bg}$). We also define $R_a$ as the artificial component of the radiance, and $R$ as the total sky radiance excluding starlight. The total radiance is defined as $R=R_a+R_{bg}$.  According to the definition of the magnitude, we can write:

\begin{equation}
    R_{bg} =  R_0 10^{-0.4 S_{bg}}
    \label{SBconv1}
\end{equation}

The zero point radiances $R_0$ are obtained from \citet{bessell1979} and given in Table \ref{zeropoints}. They were derived with the relative absolute energy distribution of \citet{hayes1970absolute} standards and the absolute flux calibration for $\alpha$ Lyrae given by \citet{hayes1975rediscussion}. Values of $R_{bg}$ and corresponding $S_{bg}$ measurements (\citet{benn1998palma} in B \& V bands and ASTMON in the R band) are given in table \ref{sbbackground} for OT in April 2019.

\begin{table}
 \centering
 \caption{Zero point radiances for the JC photometric system derived from \citet{bessell1979}. }
 \label{zeropoints}
 \begin{tabular}{cccc} 
  \hline
  Band & $\lambda_{e}$ & FWHM & $R_0$ \\
                          & nm & nm & W m$^{-2}$ sr$^{-1}$ \\
     \hline                      
        B &  436.1 & 89 & 254.3 \\
        V &  544.8 & 84 & 131.4 \\
        R &  640.7 & 158 & 151.2 \\
  \hline
 \end{tabular}
\end{table}

\begin{table}
 \centering
 \caption{Natural sky brightness, background radiances, artificial radiances and total radiances toward zenith at OT in the B V R bands. B and V were determined using \citet{benn1998palma} while R was determined with P99 measurements of table \ref{sbmeas2018-2019} .}
 \label{sbbackground}
 \begin{tabular}{ccccc} 
  \hline
Band	&	$S_{bg}$	&	$R_{bg}$	&	$R_a$	&	$R$	\\
	&	mag arcsec\textsuperscript{-2}	&	$W m^{-2} sr^{-1}$	&	$W m^{-2} sr^{-1}$	&	$W m^{-2} sr^{-1}$	\\
	\hline
B	&	22.73	&	2.06E-07	&	3.89E-08	&	2.45E-07	\\
V	&	21.93	&	2.22E-07	&	2.20E-07	&	4.42E-07	\\
R	&	21.18	&	5.10E-07	&	2.24E-07	&	7.34E-07	\\

  \hline
 \end{tabular}
\end{table}

Once $R_{bg}$ is known, the SB from any modelled artificial radiance integrated over the JC band (${R_a}$) can be determined from the definition of the magnitude.
\begin{equation}
    S = -2.5 \log \left( \frac{{R_a} + R_{bg}}{R_0} \right)
    \label{SBconv2}
\end{equation}

$R_{a}$ have to be determined by integrating the modelled artificial radiances for the spectral bins over the JC band.  
In a similar way, we can express the variation in SB ($\Delta{{S}}$) after a shutdown or after a reduction of the radiance of a given area ($\Delta{{R_a}}$). 
\begin{equation}
    \Delta{S}= -2.5 \log \left( \frac{{R_a} - \Delta{{R_a}} + R_{bg}}{{R_a} + R_{bg}} \right)
    \label{SBdelta}
\end{equation}

 Equation \ref{SBdelta} is used to calculate the values of Tables \ref{magshutdown} and \ref{magreduc}. For the case of a lamp conversion, $\Delta{{R_a}}$ is the difference in artificial radiance contribution of an area (present situation minus converted). For a complete shutdown of an area, $\Delta{{R_a}}$ is simply the present artificial radiance contribution of the area.

\section{RESULTS AND ANALYSIS}

\begin{figure}
 \includegraphics[width=\columnwidth]{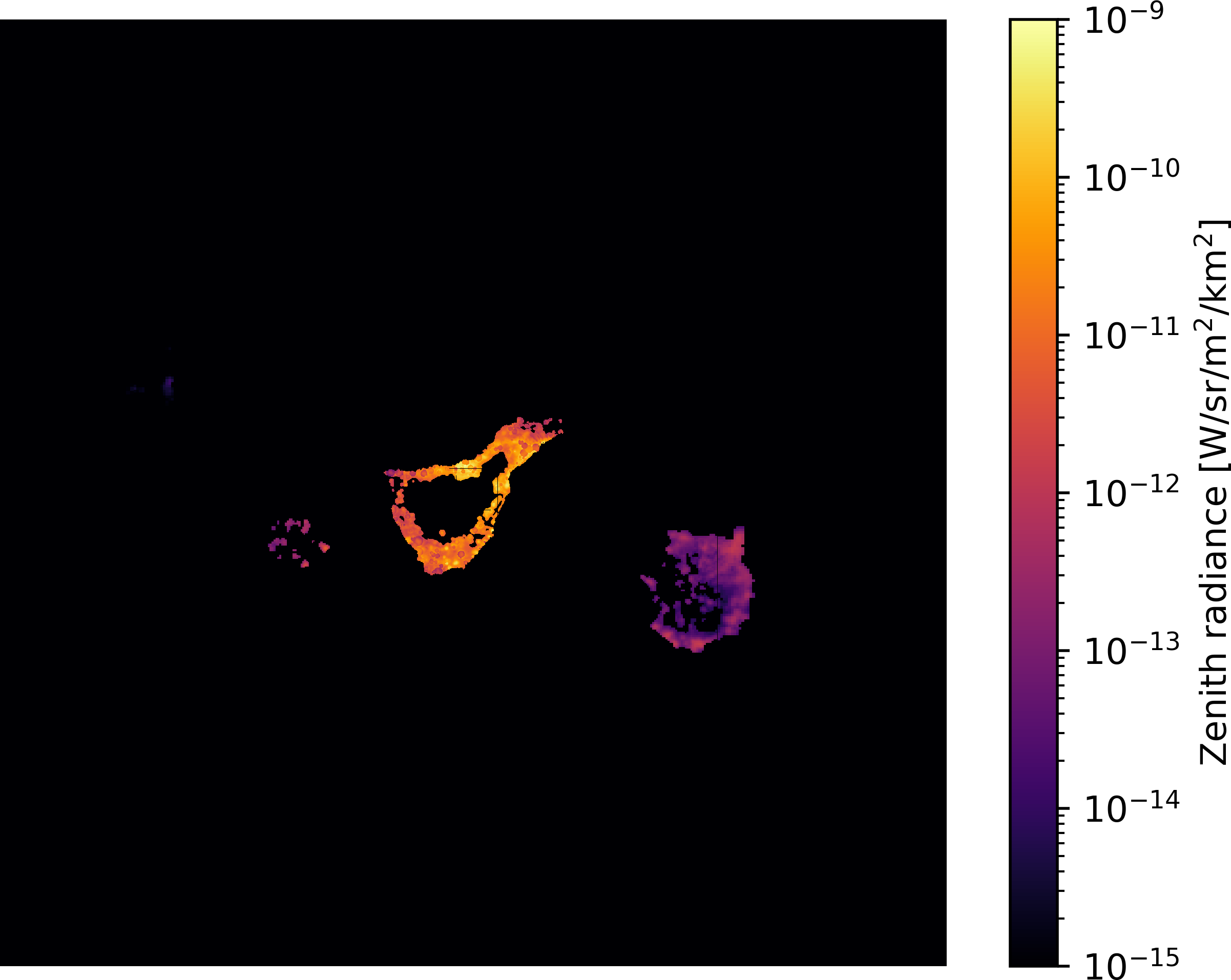}
    \caption{Contribution of the different part of the Canary Islands to the V band artificial zenith sky radiance at OT.}
    \label{contrib}
\end{figure}

\begin{figure}
 \includegraphics[width=\columnwidth]{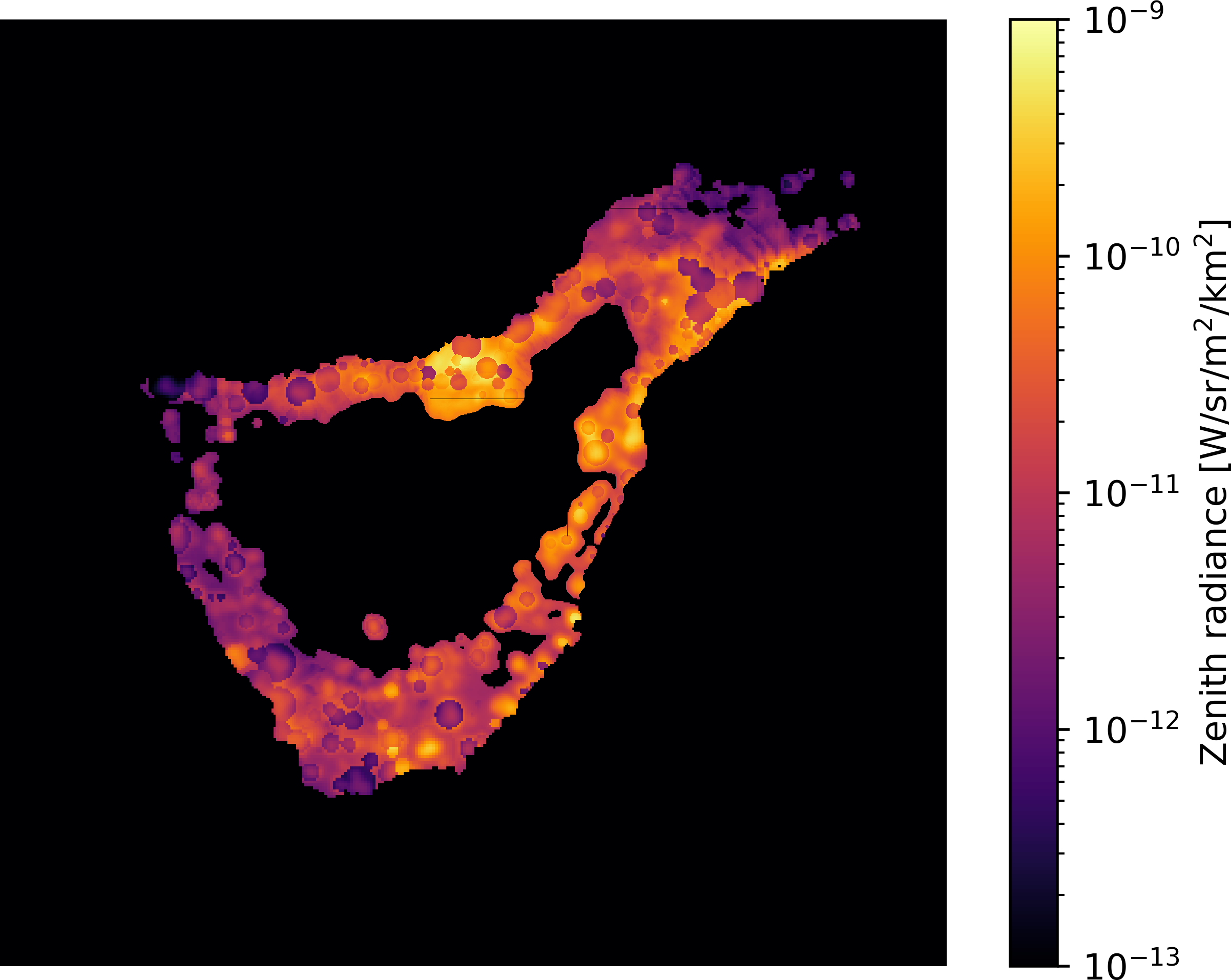}
    \caption{Contribution of the different part of the Tenerife Island to the V band artificial zenith sky radiance at OT.}
    \label{contribtnf}
\end{figure}

Figure \ref{contrib} shows the relative importance of the different part of the modelling domain in terms of their contribution to the zenith artificial radiance in the V band. On that figure, it is noticeable that most of the zenith artificial radiance at OT is coming from the island of Tenerife itself. The second contributor is Gran Canaria, and the third is the small island of La Gomera. The contribution of La Palma island is negligible. We did not calculate the effect of El Hierro but it is for sure a lot smaller. The low contribution of La Palma is easy to explain because of its modest lighting infrastructure combined to its large distance from Tenerife. Figure \ref{contribtnf} gives more details about the contribution of different parts of the Island of Tenerife. In this figure, we can clearly perceive that the contribution is a complex combination of the distance to OT and the installed lamp fluxes. It can be noticed for example that Santa Cruz de Tenerife is not a huge contributor even if it emits a large amount of light as seen from Figure \ref{viirstenerife}. A more detailed view of that information is given in Table \ref{radcontrib}. In this table, the percentage of the total artificial radiance is given for each municipality and protected/unprotected areas. This table shows that toward zenith in the V band and for the present situation, 97\% of the observed radiance comes from Tenerife Island (only 3\% comes from the other islands). The most contributing municipality is La Orotava with around 17\% followed by G\"u\'imar with around 11\%. The protected area contributes to about 43\% while it is about 54\% for the unprotected area. The capital Santa Cruz is not the most important contributor with about 7\%. Another interesting result is that Los Cristianos and Playa de Las Am\'ericas (Arona),  highest density tourists areas, contributes with about 2\%. Some of these contributions may appear low and counter intuitive. This is because that our feeling of light pollution levels on site is driven by the observation of light domes toward the main sources. Here we show that their contributions are relatively low when looking toward zenith.

In the advent of a complete conversion to LED, these numbers change a bit. The contribution of other islands becomes relatively more important (between 6\% and 11\%) but not in absolute values since in this conversion scenario only Tenerife is converted. After conversion, the relative contribution of G\"u\'imar is reduced while Los Realejos increase significantly to become the second contributing municipality. This is because that a part of Los Realejos is already converted to PCamber in the present situation, so that its absolute contribution is less reduced after the conversion in comparison to other municipalities. 

\begin{table*}
 \centering
 \caption{Fraction of the zenith artificial sky radiance to the total by municipality or area. Converted case stand for the change of all lighting devices of the protected area to PCamber with 20\% output flux reduction and the change of all lighting devices of the unprotected area to LED2700K with 70\% output flux reduction. The 5 most contributing municipalities to the present V band radiance are in bold. We also indicate their decreasing order of importance in the V band before their names.} 
 \label{radcontrib}
 \begin{tabular}{cccccccc} 
  \hline
	&		&	present	&		&		&	Fully converted	&		&		\\
Municipality / zone	&	Protected area	&	B 	&	V	&	R	&	B	&	V 	&	R	\\
	&		&	$\theta_z=0$	&	$\theta_z=0$	&	$\theta_z=0$	&	$\theta_z=0$	&	$\theta_z=0$	&	$\theta_z=0$	\\
	&		&	\%	&	\%	&	\%	&	\%	&	\%	&	\%	\\
	\hline
Adeje	&	yes	&	1.4	&	1.7	&	1.7	&	0.7	&	2.2	&	0.0	\\
Arafo	&	no	&	4.6	&	3.7	&	3.6	&	4.6	&	2.1	&	1.7	\\
Arico	&	no	&	5.4	&	6.0	&	6.1	&	6.9	&	3.2	&	2.6	\\
Arona	&	mix	&	1.8	&	2.1	&	2.2	&	2.9	&	2.2	&	2.2	\\
Buenavista	&	yes	&	0.0	&	0.0	&	0.0	&	0.0	&	0.1	&	0.1	\\
Candelaria	&	no	&	3.7	&	4.2	&	4.2	&	6.0	&	2.7	&	2.2	\\
Fasnia	&	no	&	2.3	&	2.9	&	2.9	&	4.3	&	1.9	&	1.6	\\
Garachico	&	yes	&	0.2	&	0.2	&	0.2	&	0.1	&	0.3	&	0.4	\\
\textbf{5- Granadilla}	&	no	&	6.2	&	\textbf{6.6}	&	6.7	&	8.5	&	4.1	&	3.4	\\
La Laguna	&	mix	&	3.6	&	4.5	&	4.6	&	6.5	&	5.0	&	4.9	\\
La Guancha	&	yes	&	1.0	&	1.1	&	1.1	&	0.4	&	1.3	&	1.5	\\
Guia de Isora	&	yes	&	0.5	&	0.6	&	0.6	&	0.3	&	0.8	&	0.9	\\
\textbf{2- G\"u\'imar}	&	no	&	10.8	&	\textbf{9.6}	&	9.4	&	12.8	&	5.8	&	4.7	\\
Icod de los Vinos	&	yes	&	1.0	&	1.2	&	1.2	&	0.5	&	1.5	&	1.8	\\
La Matanza	&	yes	&	0.5	&	0.9	&	0.9	&	1.0	&	1.5	&	1.6	\\
\textbf{1- La Orotava}	&	yes	&	17.4	&	\textbf{16.7}	&	16.7	&	7.0	&	22.4	&	25.5	\\
Puerto de la Cruz	&	yes	&	4.7	&	4.0	&	3.9	&	1.5	&	4.9	&	5.6	\\
\textbf{3- Los Realejos}	&	yes	&	7.0	&	\textbf{9.0}	&	9.3	&	5.7	&	13.3	&	14.9	\\
El Rosario	&	no	&	3.4	&	3.2	&	3.1	&	3.9	&	1.8	&	1.5	\\
San Juan de la Rambla	&	yes	&	1.0	&	1.2	&	1.2	&	0.5	&	1.6	&	1.8	\\
San Miguel	&	no	&	6.5	&	3.6	&	3.2	&	2.2	&	1.0	&	0.9	\\
\textbf{4- Santa Cruz de Tenerife}	&	no	&	7.7	&	\textbf{6.9}	&	6.8	&	8.5	&	3.9	&	3.2	\\
Santa Ursula	&	yes	&	1.6	&	1.9	&	2.0	&	0.9	&	2.7	&	3.1	\\
Santiago del Teide	&	yes	&	0.1	&	0.2	&	0.2	&	0.4	&	0.5	&	0.5	\\
Sauzal	&	yes	&	0.9	&	1.2	&	1.2	&	0.5	&	1.5	&	1.7	\\
Los Silos	&	yes	&	0.1	&	0.1	&	0.1	&	0.0	&	0.1	&	0.1	\\
Tacoronte	&	yes	&	0.6	&	1.4	&	1.5	&	1.3	&	2.3	&	2.6	\\
El Tanque	&	yes	&	0.2	&	0.2	&	0.2	&	0.1	&	0.2	&	0.3	\\
Tegeste	&	yes	&	0.3	&	0.4	&	0.4	&	0.2	&	0.6	&	0.6	\\
La Victoria	&	yes	&	1.3	&	1.5	&	1.5	&	0.6	&	2.0	&	2.3	\\
Vilaflor	&	no	&	0.3	&	0.3	&	0.3	&	0.3	&	0.1	&	0.1	\\
\hline
Protected zone	&	yes	&	42.5	&	46.4	&	47.1	&	27.1	&	65.2	&	73.3	\\
Unprotected zone	&	no	&	53.8	&	50.4	&	49.8	&	62.8	&	28.9	&	23.7	\\
Tenerife	&	-	&	96.3	&	96.7	&	96.9	&	89.0	&	93.7	&	93.9	\\
All	&	-	&	100.0	&	100.0	&	100.0	&	100.0	&	100.0	&	100.0	\\
\hline
Artificial radiance	&		&	3.89E-08	&	2.20E-07	&	2.24E-07	&	1.29E-08	&	1.14E-07	&	1.16E-07	\\
 W sr$^{-1}$ m$^{-2}$ 	&		&		&		&		&		&		&		\\
 \hline
 \end{tabular}
\end{table*}

\begin{figure}
 \includegraphics[width=\columnwidth]{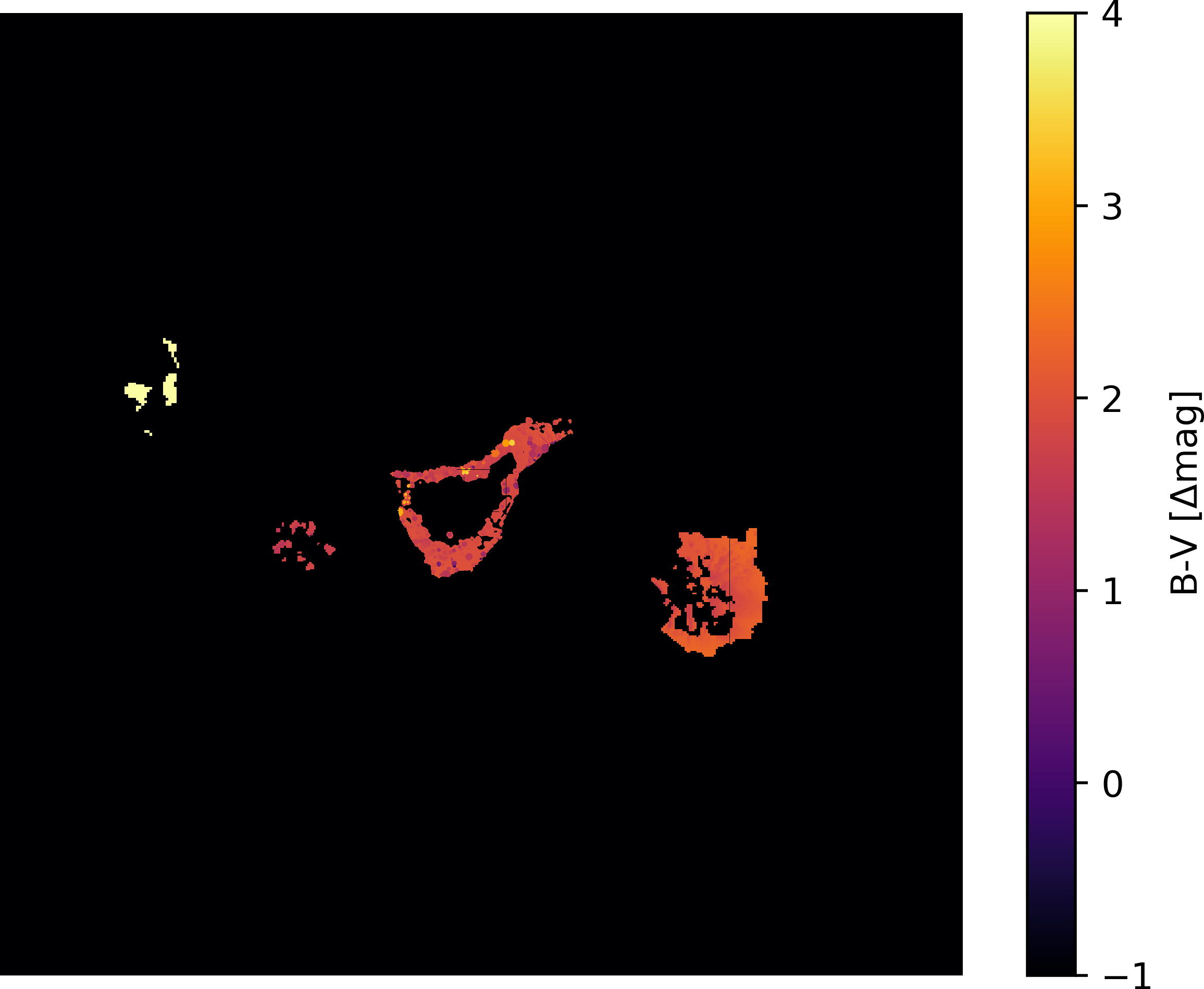}
    \caption{Colour index of the artificial sky brightness of the different parts of the Canary Islands to the artificial zenith sky radiance at OT.}
    \label{color}
\end{figure}

\begin{figure}
 \includegraphics[width=\columnwidth]{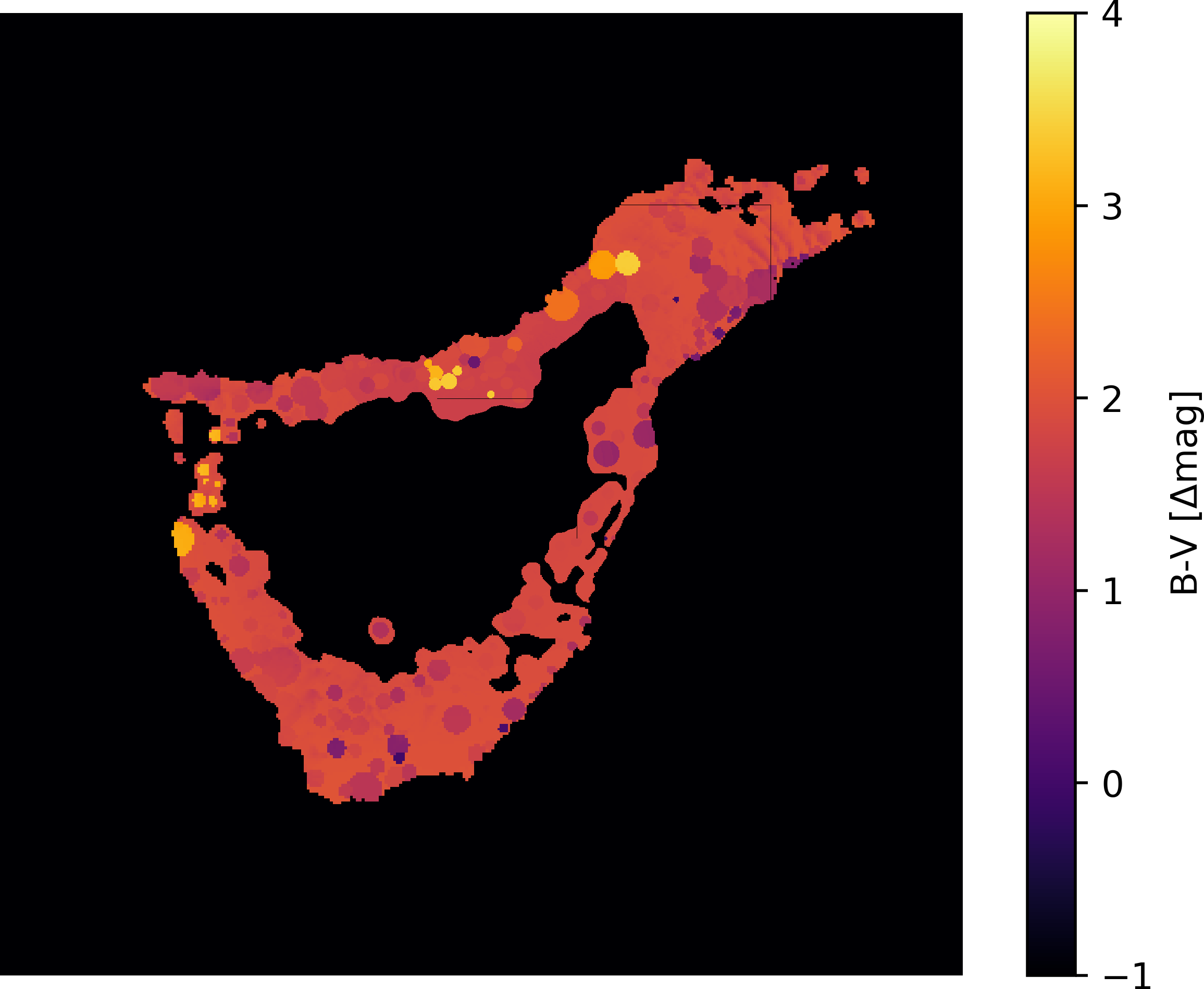}
    \caption{Colour index of the artificial sky brightness of the different parts of the Tenerife Island to the artificial zenith sky radiance at OT.}
    \label{colortnf}
\end{figure}

Figures \ref{color} and \ref{colortnf} show the $B-V$ colour index in magnitude calculated using the artificial radiances only. These figures are for the present situation. It is clear on Figure \ref{color} that La Palma does not have much blue light ($B-V \approx 4$), but there are also some low blue content spots on the island of Tenerife (Figure \ref{colortnf}), Los Realejos being one of them. These low blue radiance spots are places mostly already converted to PCamber.

Table \ref{radiancestotal} shows the total artificial radiances in B, V and R at zenith angles $\theta_z=0^\circ$ and $\theta_z=30^\circ$ for the present situation and for the conversion scenario. The zenith artificial radiance after conversion is $\approx$ 33\% of its present value in the B band while it is around 52\% in the two other filters. This result shows clearly that the blue content of the sky brightness is more efficiently reduced after conversion but the reduction is also significant in the V and R bands (about a factor of 2). These reductions are the result of a combination of the change in colour of the lamp spectra and a general reduction of the upward emitted light since the LED fixture used for the conversion have an Upward Light Output Ratio (ULOR) of 0.

\begin{table}
 \centering
 \caption{Comparison of the artificial radiances after midnight for the present situation and the full conversion.} 
 \label{radiancestotal}
 \begin{tabular}{ccccc } 
  \hline
 &  & present & Converted &  \\
Band & $\theta_z$ & Radiance & Radiance & Converted/present \\
 &  & W sr$^{-1}$ m$^{-2}$ & W sr$^{-1}$ m$^{-2}$ &  \\
 \hline
B & 0 & 3.89E-08 & 1.29E-08 & 0.33 \\
V & 0 & 2.20E-07 & 1.14E-07 & 0.52 \\
R	&	0	&	2.24E-07	&	1.16E-07	&	0.52	\\
B & 30 avg & 4.68E-08 & 1.50E-08 & 0.32 \\
V & 30 avg & 2.68E-07 & 1.36E-07 & 0.51 \\
R & 30 avg & 2.73E-07 & 1.42E-07 & 0.52 \\
  \hline
 \end{tabular}
\end{table}

One interesting aspect of modelling the sky brightness is that we can a test unlimited number of changes in the lighting infrastructure and environmental variables to determine their effects on the sky brightness. Table \ref{magshutdown} shows the SB change associated with the shutdown of each municipality. This table shows that a complete shutdown of Tenerife Island would improve zenith OT SB by 0.155 mag arcsec\textsuperscript{-2} in the B band, 0.427 mag arcsec\textsuperscript{-2} in the V band and 0.258 mag arcsec\textsuperscript{-2} in the R band. These numbers are the maximal SB reductions availables but implies a complete shutdown of the light fixtures. That is certainly not realistic in the real world. With such a SB reduction, OT sky would be as dark as ORM sky. The five most contributing municipalities to the zenith SB in V band are in order of decreasing importance: La Orotava, G\"u\'imar, Los Realejos, Santa Cruz de Tenerife and Granadilla. The SB reductions at $\theta_z=30^\circ$ are even larger (e.g., 0.461 mag arcsec\textsuperscript{-2} in the V band). 

\begin{table*}
 \centering
 \caption{Reduction of the sky brightness after midnight in the B V R JC bands if a municipality or area is totally shutdown. The 5 most contributing municipalities to the present V band SB are in bold. We also indicate their decreasing order of importance in the V band before their names.
 }
 \label{magshutdown}
 \begin{tabular}{cccccccc}
  \hline
Municipality / zone	&	Protected area	&	B 	&	V  	&	R  	&	B  	&	V  	&	R  \\
	&		&	$\theta_z=0$	&	$\theta_z=0$	&	$\theta_z=0$	&	$\theta_z=30^o$	&	$\theta_z=30^o$	&	$\theta_z=30^o$	\\
	&		&	mag arcsec\textsuperscript{-2}	&	mag arcsec\textsuperscript{-2}	&	mag arcsec\textsuperscript{-2}	&	mag arcsec\textsuperscript{-2}	&	mag arcsec\textsuperscript{-2}	&	mag arcsec\textsuperscript{-2}	\\
	\hline
Adeje	&	yes	&	0.002	&	0.009	&	0.006	&	0.003	&	0.010	&	0.007	\\
Arafo	&	no	&	0.008	&	0.020	&	0.012	&	0.009	&	0.022	&	0.013	\\
Arico	&	no	&	0.009	&	0.032	&	0.020	&	0.011	&	0.035	&	0.023	\\
Arona	&	mix	&	0.003	&	0.011	&	0.007	&	0.004	&	0.013	&	0.008	\\
Buenavista	&	yes	&	0.000	&	0.000	&	0.000	&	0.000	&	0.000	&	0.000	\\
Candelaria	&	no	&	0.006	&	0.022	&	0.014	&	0.007	&	0.024	&	0.016	\\
Fasnia	&	no	&	0.004	&	0.015	&	0.010	&	0.005	&	0.016	&	0.011	\\
Garachico	&	yes	&	0.000	&	0.001	&	0.001	&	0.000	&	0.001	&	0.001	\\
\textbf{5- Granadilla}	&	no	&	0.011	&	\textbf{0.035}	&	0.022	&	0.013	&	0.040	&	0.026	\\
La Laguna	&	mix	&	0.006	&	0.024	&	0.015	&	0.007	&	0.027	&	0.018	\\
La Guancha	&	yes	&	0.002	&	0.006	&	0.004	&	0.002	&	0.006	&	0.004	\\
Guia de Isora	&	yes	&	0.001	&	0.003	&	0.002	&	0.001	&	0.003	&	0.002	\\
\textbf{2- G\"u\'imar}	&	no	&	0.018	&	\textbf{0.051}	&	0.031	&	0.021	&	0.054	&	0.034	\\
Icod de los Vinos	&	yes	&	0.002	&	0.006	&	0.004	&	0.002	&	0.007	&	0.005	\\
La Matanza	&	yes	&	0.001	&	0.005	&	0.003	&	0.001	&	0.005	&	0.003	\\
\textbf{1- La Orotava}	&	yes	&	0.030	&	\textbf{0.087}	&	0.054	&	0.034	&	0.092	&	0.059	\\
Puerto de la Cruz	&	yes	&	0.008	&	0.021	&	0.013	&	0.009	&	0.023	&	0.015	\\
\textbf{3- Los Realejos}	&	yes	&	0.012	&	\textbf{0.048}	&	0.030	&	0.013	&	0.050	&	0.033	\\
El Rosario	&	no	&	0.006	&	0.017	&	0.010	&	0.007	&	0.019	&	0.012	\\
San Juan de la Rambla	&	yes	&	0.002	&	0.006	&	0.004	&	0.002	&	0.007	&	0.004	\\
San Miguel	&	no	&	0.011	&	0.019	&	0.011	&	0.013	&	0.022	&	0.013	\\
\textbf{4- Santa Cruz de Tenerife}	&	no	&	0.013	&	\textbf{0.037}	&	0.022	&	0.016	&	0.042	&	0.027	\\
Santa Ursula	&	yes	&	0.003	&	0.010	&	0.007	&	0.003	&	0.012	&	0.008	\\
Santiago del Teide	&	yes	&	0.000	&	0.001	&	0.001	&	0.000	&	0.001	&	0.001	\\
Sauzal	&	yes	&	0.002	&	0.006	&	0.004	&	0.002	&	0.007	&	0.005	\\
Los Silos	&	yes	&	0.000	&	0.000	&	0.000	&	0.000	&	0.001	&	0.000	\\
Tacoronte	&	yes	&	0.001	&	0.007	&	0.005	&	0.001	&	0.008	&	0.006	\\
El Tanque	&	yes	&	0.000	&	0.001	&	0.001	&	0.000	&	0.001	&	0.001	\\
Tegeste	&	yes	&	0.000	&	0.002	&	0.001	&	0.001	&	0.002	&	0.001	\\
La Victoria	&	yes	&	0.002	&	0.008	&	0.005	&	0.003	&	0.009	&	0.006	\\
Vilaflor	&	no	&	0.000	&	0.002	&	0.001	&	0.001	&	0.002	&	0.001	\\
\hline
Protected zone	&	yes	&	0.071	&	0.226	&	0.146	&	0.081	&	0.243	&	0.163	\\
Unprotected zone	&	no	&	0.089	&	0.243	&	0.154	&	0.104	&	0.266	&	0.175	\\
Tenerife	&	-	&	0.155	&	0.427	&	0.281	&	0.178	&	0.461	&	0.316	\\
All	&	-	&	0.160	&	0.439	&	0.289	&	0.184	&	0.474	&	0.325	\\
  \hline 
 \end{tabular}
\end{table*}

A more realistic scenario is presented in Tables \ref{magreduc} and \ref{radredpop}. Table \ref{magreduc} shows the expected SB reduction after a conversion of a municipality or area to the relevant LED technology. As shown in the table, no gain in zenith SB may be achieved with the lighting conversion of other islands to PCamber. Furthermore, only a tiny reduction ($\leq 0.002$ mag arcsec\textsuperscript{-2}) can be obtained at $30^\circ$ zenith angle from the conversions of the other islands. The five municipalities conversion that may deliver the maximum zenith V band SB reduction are, in order of decreasing effect, 1- G\"u\'imar (pop. 20\,190), 2- La Orotava (pop. 42\,029), 3- Santa Cruz de Tenerife (pop. 207\,312), 4- Granadilla (pop. 50\,146), 5- Arico (pop. 7\,988). The available SB reduction is relatively similar from one to the other municipality with 0.036 mag arcsec\textsuperscript{-2} for G\"u\'imar and 0.025 mag arcsec\textsuperscript{-2} for Arico. Some of these 5 municipalities are clearly less populated and thus involve less light points to be converted. These municipalities should be prioritized to get the maximum SB reduction with minimal investment. The ratio of total sky radiance reduction per inhabitant for each municipality is shown in Table \ref{radredpop}. If we consider the maximum zenith sky radiance reduction for minimal investment in V band, Fasnia should certainly be the best starting point followed by Arico, Arafo, G\"u\'imar and San Miguel. These five municipalities may reduce the total V band sky radiance by 9.2\% with only 6.25 \% of the Tenerife population, compared to the complete conversion o the island that should result in a radiance reduction of 24.1\%. In terms of SB, the complete conversion of the Tenerife Island should improve the zenith V band SB by 0.299 mag arcsec\textsuperscript{-2}. This is actually the best SB reduction in the V band that can be achieved with the lighting conversion rules currently in place. The conversion of the five municipalities listed above should deliver a SB reduction of $\approx$ 0.1 mag arcsec\textsuperscript{-2} in the V band.

\begin{table*}
 \centering
 \caption{Reduction of the sky brightness after midnight in the B V R JC bands if a municipality or area is converted to LED. PCamber with 20\% output flux reduction in the protected area and LED2700K with 70\% output flux reduction in the unprotected area. Output flux assumed to remain constant for other islands. The 5 municipalities with the largest reduction of the V band SB after a conversion to LED are in bold. We also indicate their decreasing order of importance in the V band before their names. Note that the ordre is different from tables \ref{radcontrib} and \ref{magshutdown}.
 }
 \label{magreduc}
 \begin{tabular}{ccccccccc} 
  \hline
Municipality / zone	&	Protected area	&	B	&	V 	&	R 	&	B 	&	V 	&	R 	\\
	&		&	$\theta_z=0$	&	$\theta_z=0$	&	$\theta_z=0$	&	$\theta_z=30^o$	&	$\theta_z=30^o$	&	$\theta_z=30^o$	\\
	&		&	mag arcsec\textsuperscript{-2}	&	mag arcsec\textsuperscript{-2}	&	mag arcsec\textsuperscript{-2}	&	mag arcsec\textsuperscript{-2}	&	mag arcsec\textsuperscript{-2}	&	mag arcsec\textsuperscript{-2}	\\
	\hline
Adeje	&	yes	&	0.002	&	0.003	&	0.006	&	0.002	&	0.002	&	0.001	\\
Arafo	&	no	&	0.005	&	0.014	&	0.009	&	0.005	&	0.010	&	0.008	\\
\textbf{5- Arico}	&	no	&	0.005	&	\textbf{0.024}	&	0.016	&	0.005	&	0.017	&	0.013	\\
Arona	&	mix	&	0.001	&	0.005	&	0.003	&	0.002	&	0.004	&	0.003	\\
Buenavista	&	yes	&	0.000	&	0.000	&	0.000	&	0.000	&	0.000	&	0.000	\\
Candelaria	&	no	&	0.003	&	0.015	&	0.010	&	0.003	&	0.010	&	0.008	\\
Fasnia	&	no	&	0.002	&	0.010	&	0.007	&	0.002	&	0.007	&	0.006	\\
Garachico	&	yes	&	0.000	&	0.000	&	0.000	&	0.000	&	0.000	&	0.000	\\
\textbf{4- Granadilla}	&	no	&	0.006	&	\textbf{0.025}	&	0.016	&	0.006	&	0.018	&	0.014	\\
La Laguna	&	mix	&	0.002	&	0.010	&	0.007	&	0.004	&	0.008	&	0.006	\\
La Guancha	&	yes	&	0.001	&	0.002	&	0.001	&	0.001	&	0.002	&	0.001	\\
Guia de Isora	&	yes	&	0.001	&	0.001	&	0.000	&	0.001	&	0.001	&	0.000	\\
\textbf{1- G\"u\'imar}	&	no	&	0.011	&	\textbf{0.036}	&	0.023	&	0.011	&	0.025	&	0.019	\\
Icod de los Vinos	&	yes	&	0.001	&	0.002	&	0.001	&	0.002	&	0.001	&	0.001	\\
La Matanza	&	yes	&	0.000	&	0.001	&	0.000	&	0.000	&	0.000	&	0.000	\\
\textbf{2- La Orotava}	&	yes	&	0.026	&	\textbf{0.028}	&	0.011	&	0.025	&	0.019	&	0.010	\\
Puerto de la Cruz	&	yes	&	0.007	&	0.008	&	0.003	&	0.007	&	0.006	&	0.003	\\
Los Realejos	&	yes	&	0.009	&	0.012	&	0.005	&	0.008	&	0.008	&	0.004	\\
El Rosario	&	no	&	0.004	&	0.012	&	0.008	&	0.004	&	0.009	&	0.007	\\
San Juan de la Rambla	&	yes	&	0.002	&	0.002	&	0.001	&	0.001	&	0.001	&	0.001	\\
San Miguel	&	no	&	0.010	&	0.017	&	0.009	&	0.010	&	0.012	&	0.008	\\
\textbf{3- Santa Cruz de Tenerife}	&	no	&	0.008	&	\textbf{0.027}	&	0.017	&	0.009	&	0.020	&	0.015	\\
Santa Ursula	&	yes	&	0.002	&	0.003	&	0.001	&	0.002	&	0.002	&	0.001	\\
Santiago del Teide	&	yes	&	0.000	&	0.000	&	0.000	&	0.000	&	0.000	&	0.000	\\
Sauzal	&	yes	&	0.001	&	0.002	&	0.001	&	0.001	&	0.001	&	0.001	\\
Los Silos	&	yes	&	0.000	&	0.000	&	0.000	&	0.000	&	0.000	&	0.000	\\
Tacoronte	&	yes	&	0.000	&	0.001	&	0.000	&	0.000	&	0.001	&	0.000	\\
El Tanque	&	yes	&	0.000	&	0.001	&	0.000	&	0.000	&	0.000	&	0.000	\\
Tegeste	&	yes	&	0.000	&	0.000	&	0.000	&	0.000	&	0.000	&	0.000	\\
La Victoria	&	yes	&	0.002	&	0.003	&	0.001	&	0.002	&	0.002	&	0.001	\\
Vilaflor	&	no	&	0.000	&	0.001	&	0.001	&	0.000	&	0.001	&	0.001	\\
\hline
Protected zone	&	yes	&	0.059	&	0.071	&	0.030	&	0.058	&	0.050	&	0.025	\\
Unprotected zone	&	no	&	0.058	&	0.211	&	0.132	&	0.058	&	0.147	&	0.112	\\
Tenerife	&	-	&	0.122	&	0.299	&	0.172	&	0.121	&	0.205	&	0.141	\\
All	&	-	&	0.122	&	0.299	&	0.172	&	0.122	&	0.208	&	0.144	\\
  \hline
 \end{tabular}
\end{table*}

\begin{table*}
 \centering
 \caption{Relative reduction of the total zenith sky radiances (artificial + natural) in the B V R bands and equivalent reduction per inhabitant \citep{population}. The municipalities in bold are the 5 most efficient to reduce the zenith V band sky brightness when considering the investment per inhabitant. We estimate that the replacement program should focus first on that list of municipalities.
 }
 \label{radredpop}
 \begin{tabular}{ccccccccc}
  \hline
	&		&	Relative	&		&		&	Relative	&		&		\\
	&		&	reduction	&		&		&	reduction	&		&		\\
	&		&		&		&		&	per inhabitant	&		&		\\
Municipality / zone	&	Population	&	B 	&	V 	&	R	&	B 	&	V 	&	R 	\\
	&		&	$\theta_z=0$	&	$\theta_z=0$	&	$\theta_z=0$	&	$\theta_z=0$	&	$\theta_z=0$	&	$\theta_z=0$	\\
	&		&	\%	&	\%	&	\%	&	\% per inhabitant $\times 10^{6}$	&	\% per inhabitant $\times 10^{6}$	&	\% per inhabitant $\times 10^{6}$	\\
	\hline
Adeje	&	47869	&	0.19	&	0.3	&	0.5	&	4.0	&	5.3	&	10.8	\\
\textbf{3- Arafo}	&	5551	&	0.49	&	1.3	&	0.8	&	87.9	&	\textbf{237.5}	&	149.6	\\
\textbf{2- Arico}	&	7988	&	0.49	&	2.1	&	1.4	&	61.5	&	\textbf{268.6}	&	179.8	\\
Arona	&	81216	&	0.13	&	0.5	&	0.3	&	1.6	&	6.0	&	3.9	\\
Buenavista	&	4778	&	0.00	&	0.0	&	0.0	&	1.0	&	1.0	&	0.4	\\
Candelaria	&	27985	&	0.27	&	1.4	&	0.9	&	9.7	&	49.0	&	33.2	\\
\textbf{1- Fasnia}	&	2786	&	0.15	&	0.9	&	0.6	&	52.1	&	\textbf{333.7}	&	230.5	\\
Garachico	&	4871	&	0.02	&	0.0	&	0.0	&	4.6	&	5.3	&	2.1	\\
Granadilla	&	50146	&	0.54	&	2.3	&	1.5	&	10.8	&	45.0	&	29.9	\\
La Laguna	&	157503	&	0.23	&	0.9	&	0.6	&	1.4	&	5.9	&	3.9	\\
La Guancha	&	5520	&	0.14	&	0.2	&	0.1	&	24.8	&	36.2	&	17.9	\\
Guia de Isora	&	21368	&	0.06	&	0.1	&	0.0	&	2.8	&	3.4	&	1.4	\\
\textbf{4- G\"u\'imar}	&	20190	&	1.03	&	3.3	&	2.1	&	51.2	&	\textbf{162.4}	&	105.4	\\
Icod de los Vinos	&	23254	&	0.14	&	0.2	&	0.1	&	5.9	&	7.9	&	3.6	\\
La Matanza	&	9061	&	0.03	&	0.0	&	0.0	&	3.5	&	5.4	&	2.7	\\
La Orotava	&	42029	&	2.40	&	2.5	&	1.1	&	57.1	&	60.6	&	25.0	\\
Puerto de la Cruz	&	30468	&	0.67	&	0.7	&	0.3	&	22.0	&	23.4	&	9.8	\\
Los Realejos	&	36402	&	0.81	&	1.1	&	0.5	&	22.3	&	29.0	&	13.0	\\
El Rosario	&	17370	&	0.34	&	1.1	&	0.7	&	19.5	&	64.3	&	41.6	\\
San Juan de la Rambla	&	4828	&	0.14	&	0.2	&	0.1	&	28.8	&	36.7	&	16.5	\\
\textbf{5- San Miguel}	&	20886	&	0.92	&	1.5	&	0.9	&	44.2	&	\textbf{73.3}	&	40.8	\\
Santa Cruz de Tenerife	&	207312	&	0.77	&	2.4	&	1.6	&	3.7	&	11.8	&	7.5	\\
Santa Ursula	&	14679	&	0.21	&	0.3	&	0.1	&	14.5	&	18.0	&	8.0	\\
Santiago del Teide	&	11111	&	0.00	&	0.0	&	0.0	&	0.0	&	0.0	&	0.0	\\
Sauzal	&	8934	&	0.12	&	0.2	&	0.1	&	13.9	&	20.9	&	10.3	\\
Los Silos	&	4693	&	0.01	&	0.0	&	0.0	&	2.2	&	3.2	&	1.6	\\
Tacoronte	&	24134	&	0.03	&	0.1	&	0.0	&	1.4	&	3.1	&	1.7	\\
El Tanque	&	2763	&	0.03	&	0.1	&	0.0	&	11.5	&	20.0	&	10.7	\\
Tegeste	&	11294	&	0.04	&	0.0	&	0.0	&	3.2	&	3.5	&	1.3	\\
La Victoria	&	9185	&	0.17	&	0.2	&	0.1	&	19.0	&	25.7	&	12.1	\\
Vilaflor	&	1667	&	0.03	&	0.1	&	0.1	&	16.3	&	64.4	&	42.7	\\
\hline
Protected zone	&	-	&	5.32	&	6.3	&	2.8	&	-	&	-	&	-	\\
Unprotected zone	&	-	&	5.24	&	17.6	&	11.5	&	-	&	-	&	-	\\
Tenerife	&	917841	&	10.61	&	24.1	&	14.7	&	11.6	&	26.2	&	16.0	\\
All	&	-	&	10.61	&	24.1	&	14.7	&	-	&	-	&	-	\\
\hline
Total radiance	&		&	2.45E-07	&	4.42E-07	&	7.34E-07	&	-	&	-	&	-	\\
$W m^{-2} sr^{-1}$	&		&		&		&		&		&		&		\\
  \hline
 \end{tabular}
\end{table*}

\section{Conclusions}

In this paper we show how a radiative transfer code dedicated to the modelling of the sky radiance can be used to plan an efficient light conversion to restore the night sky brightness to ist natural value. The methodology presented is applied to the case of Observatorio del Teide, Tenerife. We showed how the determination of the sky radiance for the present situation and for some LED conversion plans can be combined to optimize the night sky darkness restoration. The integration of the results according to the Johnson-Cousins bands spectral responses and for defined territories (like municipality limits) render it possible to identify the most urgent municipalities to convert in order to get the maximum decrease of the sky brightness with the minimum financial and human resources. 

We demonstrated that just the completion of the undergoing lighting infrastructure conversion plan of the Tenerife Island should translate into a V band sky brightness reduction of $\approx$ 0.3 mag arcsec\textsuperscript{-2}. Such improvement would not be enough to recover a sky darkness typical  of astronominal sites like Observatorio del Roque de los Muchachos, however it is not that far from it. We would need a zenith V band sky brightness reduction of $\approx$ 0.44 mag arcsec\textsuperscript{-2} to be reaching the same sky darkness. A reduction of $\approx$ 0.3 mag arcsec\textsuperscript{-2} means a reduction of the total V band sky radiance by 24\% and a reduction of the artificial V band sky radiance by 48\%. This is actually a dramatic reduction of the light pollution. A complete shutdown of the Tenerife light would improve the V band zenith sky brightness by $\approx$ 0.43 mag arcsec\textsuperscript{-2}. 

The capital Santa Cruz de Tenerife and the busy tourist places of Los Cristianos and Playa de Las Américas (Arona) are producing respectively 7\% and 2\% of the artificial sky radiance toward zenith. Their contribution would be dominant when looking closer to horizon in their respective directions but it was not evaluated in this study. Actually, our results show clearly that nearby sources are the main contributors to the sky brightness toward zenith. We can expect that, for some specific research applications needing to observe closer to horizon, it can represents an important problem and then any further touristic development should consider strong mitigation measures to restrict their light pollution emissions.

We also showed that the sky radiance reduction per inhabitant can be an efficient proxy to optimize  sky brightness reductions with limited resources. Applying the conversion plan for the 5 municipalities showing the highest ratio of radiance reduction per inhabitant (see table \ref{radredpop}) allows the reduction of the sky brightness by $\approx$ 0.1 mag arcsec\textsuperscript{-2} (i.e., a total V band sky radiance reduction of $\approx$ 9\% compared to the maximum available of $\approx$ 24\%). But these municipalities represent only about 6\% of the Tenerife population.

This study showed how the atmospheric correction and, most importantly, the obstacles blocking correction play significant role in determining the lamp fluxes from the VIIRS-DNB radiances. For that reason, we will emphasize the addition of this feature to the Illumina model.

\section*{Acknowledgements}

We applied the sequence-determines-credit approach \citep{tscharntke2007author} for the sequence of authors. An important part of that research funded by M. Aubé's Fonds de recherche du Québec -- Nature et technologies grant. Most computations were carried out on the Mammouth Parallel II cluster managed by Calcul Québec and Compute Canada. The operation of these supercomputers is funded by the Canada Foundation for Innovation (CFI), NanoQuébec, Réseau de Médecine Génétique Appliquée, and the Fonds de recherche du Québec -- Nature et technologies (FRQNT). Thanks to Julio A. Castro Almazán for his help in processing the ASTMON data. This study was carried out in part at IAC, we want to thank that institution and its deputy director for the financial support and warm welcome.




\bibliographystyle{mnras}
\bibliography{example} 


\newpage
\appendix
\onecolumn

\begin{landscape}
\section{Light fixture and obstacles inventory for the present situation}

\begin{longtable}{lccccccl}
\caption{Light fixture and obstacles inventory} \\
\hline \multicolumn{1}{l}{\textbf{Latitude}} & \multicolumn{1}{c}{\textbf{Longitude}} & \multicolumn{1}{c}{\textbf{Radius}} & \multicolumn{1}{c}{\textbf{Obst. height}} & \multicolumn{1}{c}{\textbf{Obst. Distance}} & \multicolumn{1}{c}{\textbf{Obst. Filling factor}} & \multicolumn{1}{c}{\textbf{Lamp height}} & \multicolumn{1}{l}{\textbf{Lamp spectra and ULOR}} \\ 
degree & degree & km & m & m & - & m &          \\
\hline
\endfirsthead
\multicolumn{8}{c}{\tablename\ \thetable{} -- continued from previous page} \\
\hline \multicolumn{1}{l}{\textbf{Latitude}} & \multicolumn{1}{c}{\textbf{Longitude}} & \multicolumn{1}{c}{\textbf{Radius}} & \multicolumn{1}{c}{\textbf{Obst. height}} & \multicolumn{1}{c}{\textbf{Obst. Distance}} & \multicolumn{1}{c}{\textbf{Obst. Filling factor}} & \multicolumn{1}{c}{\textbf{Lamp height}} & \multicolumn{1}{l}{\textbf{Lamp spectra and ULOR}} \\ 
degree & degree & km & m & m & - & m &          \\
\hline
\endhead
\hline \hline
\endlastfoot
28.30094 & -16.510816 & 49.97 & 5 & 7 & 0.2 & 6 & 100\_HPS\_0   \\
27.93533 & -15.598841 & 35.15 & 15 & 15 & 0.9 & 6 & 90\_HPS\_2 10\_L40\_0  \\
28.66761 & -17.848418 & 25.76 & 15 & 12 & 0.95 & 6 & 15\_PCA\_0 10\_HPS\_0 75\_LPS\_0 \\
28.46752 & -16.592312 & 19.5 & 6 & 10 & 0.2 & 5 & 100\_HPS\_2   \\
28.24940 & -16.814967 & 19.27 & 8 & 35 & 0.1 & 9 & 100\_HPS\_0   \\
28.63933 & -16.359244 & 17.11 & 8 & 35 & 0.1 & 9 & 100\_HPS\_0   \\
28.12460 & -17.230469 & 14.59 & 7 & 15 & 0.8 & 6 & 90\_HPS\_2 10\_L40\_0  \\
28.33264 & -16.396496 & 5.07 & 5 & 100 & 0.1 & 9 & 100\_HPS\_0   \\
28.04566 & -16.575587 & 2.44 & 13 & 100 & 0.25 & 9 & 100\_HPS\_0   \\
28.51875 & -16.387880 & 2.08 & 6 & 30 & 0.4 & 7 & 99\_HPS\_0 1\_PCA\_0 \\ 
28.12210 & -16.734982 & 1.97 & 12 & 18 & 0.7 & 9 & 100\_HPS\_0   \\
28.37376 & -16.851106 & 1.95 & 10 & 10 & 0.9 & 6 & 100\_HPS\_0   \\
28.08387 & -16.732077 & 1.79 & 16 & 28 & 0.6 & 8 & 10\_PCA\_0 2\_L27\_0 78\_HPS\_0 \\
28.01340 & -16.649815 & 1.73 & 14 & 12 & 0.9 & 6 & 80\_HPS\_0 15\_MV3\_0 5\_PCA\_0 \\
28.51767 & -16.300309 & 1.723 & 6 & 25 & 0.3 & 6 & 100\_HPS\_0   \\
28.44829 & -16.458228 & 1.71 & 8 & 30 & 0.5 & 7 & 50\_PCA\_0 50\_HPS\_1  \\
28.23371 & -16.841736 & 1.69 & 16 & 22 & 0.7 & 7 & 97\_PCA\_0 3\_L27\_0  \\
28.46672 & -16.256862 & 1.68 & 24 & 30 & 0.9 & 9 & 90\_HPS\_0 5\_L40\_0 5\_MH\_1 \\
28.42892 & -16.493302 & 1.67 & 24 & 30 & 0.7 & 6 & 100\_HPS\_0   \\
28.10089 & -16.755249 & 1.67 & 12 & 30 & 0.6 & 9 & 100\_HPS\_0   \\
28.37183 & -16.815418 & 1.62 & 8 & 10 & 0.9 & 4 & 100\_HPS\_3   \\
28.44793 & -16.305677 & 1.59 & 9 & 11 & 0.9 & 5 & 95\_HPS\_0 5\_MH\_1  \\
28.46175 & -16.285635 & 1.56 & 32 & 16 & 0.75 & 6 & 95\_HPS\_0 5\_MH\_1  \\
28.39869 & -16.572359 & 1.49 & 5 & 30 & 0.8 & 4 & 100\_HPS\_0   \\
28.41183 & -16.545088 & 1.49 & 20 & 15 & 0.8 & 7 & 60\_PCA\_0 21\_L27\_0 19\_HPS\_0 \\
28.36733 & -16.714263 & 1.47 & 24 & 7 & 0.9 & 4 & 100\_HPS\_0   \\
28.48344 & -16.416816 & 1.45 & 8 & 30 & 0.5 & 3 & 80\_PCA\_0 20\_HPS\_0  \\
28.07720 & -16.557724 & 1.41 & 8 & 10 & 0.9 & 7 & 100\_HPS\_0   \\
28.46759 & -16.379026 & 1.38 & 8 & 40 & 0.4 & 9 & 100\_HPS\_0   \\
28.33341 & -16.370738 & 1.38 & 7 & 50 & 0.85 & 10 & 90\_HPS\_0 2\_L40\_0 8\_MH\_5 \\
28.37304 & -16.785446 & 1.37 & 8 & 25 & 0.4 & 8 & 100\_HPS\_0   \\
28.02369 & -16.615692 & 1.33 & 10 & 26 & 0.7 & 5 & 100\_HPS\_0   \\
28.05784 & -16.731080 & 1.32 & 14 & 40 & 0.75 & 9 & 10\_PCA\_0 2\_L27\_0 78\_HPS\_0 \\
28.36792 & -16.760711 & 1.32 & 7 & 6 & 0.9 & 6 & 50\_HPS\_0 50\_HPS\_1  \\
28.48450 & -16.341955 & 1.31 & 5 & 60 & 0.5 & 9 & 100\_HPS\_0   \\
28.31567 & -16.411176 & 1.31 & 10 & 16 & 0.85 & 6 & 95\_HPS\_0 3\_HPS\_5 2\_MH\_20 \\
28.12714 & -16.774962 & 1.3 & 5 & 25 & 0.75 & 4 & 25\_HPS\_3 5\_HPS\_50 70\_HPS\_0 \\
28.47380 & -16.304124 & 1.28 & 12 & 10 & 0.9 & 9 & 95\_HPS\_0 5\_MH\_1  \\
28.37888 & -16.686270 & 1.28 & 9 & 15 & 0.7 & 5 & 100\_HPS\_0   \\
28.48549 & -16.391648 & 1.18 & 9 & 17 & 0.5 & 5 & 100\_PCA\_0   \\
28.16630 & -16.501999 & 1.16 & 5 & 8 & 0.8 & 7 & 90\_HPS\_0 10\_HPS\_1  \\
28.44652 & -16.264041 & 1.16 & 5 & 18 & 0.8 & 7 & 95\_HPS\_2 2.5\_MH\_2 2.5\_FLU\_2 \\
28.05080 & -16.711846 & 1.15 & 16 & 60 & 0.6 & 9 & 10\_PCA\_0 2\_L27\_0 78\_HPS\_0 \\
28.08682 & -16.500563 & 1.14 & 8 & 65 & 0.3 & 9 & 5\_MH\_10 95\_HPS\_0  \\
28.52364 & -16.344300 & 1.14 & 7 & 12 & 0.7 & 7 & 100\_HPS\_0   \\
28.35083 & -16.703138 & 1.12 & 7 & 10 & 0.5 & 7 & 50\_HPS\_3 50\_HPS\_0  \\
28.05334 & -16.616540 & 1.1 & 8 & 35 & 0.6 & 9 & 5\_MH\_10 95\_HPS\_0  \\
28.56943 & -16.324811 & 1.1 & 8 & 6 & 0.6 & 7 & 95\_HPS\_0 3\_PCA\_0 2\_L27\_0 \\
28.39099 & -16.523835 & 1.09 & 9 & 12 & 0.9 & 5 & 100\_HPS\_0   \\
28.48621 & -16.318925 & 1.08 & 18 & 12 & 0.9 & 9 & 75\_HPS\_0 10\_L40\_0 10\_L30\_0 \\
28.12130 & -16.576724 & 1.08 & 10 & 20 & 0.9 & 6 & 3\_HPS\_38 97\_HPS\_0  \\
28.21147 & -16.778658 & 1.07 & 8 & 5 & 0.9 & 4 & 50\_HPS\_0 50\_HPS\_3  \\
28.50047 & -16.317123 & 1.05 & 7 & 12 & 0.6 & 5 & 90\_HPS\_0 10\_L40\_0  \\
28.36897 & -16.368080 & 1.03 & 8 & 30 & 0.8 & 6 & 100\_HPS\_0   \\
28.46480 & -16.403630 & 1.03 & 9 & 15 & 0.7 & 6 & 100\_HPS\_0   \\
28.49336 & -16.201951 & 0.98 & 6 & 35 & 0.4 & 8 & 95\_HPS\_0 2.5\_MH\_2 2.5\_FLU\_2 \\
28.04623 & -16.537383 & 0.97 & 20 & 15 & 0.8 & 7 & 100\_HPS\_0   \\
28.07029 & -16.655582 & 0.97 & 8 & 15 & 0.75 & 9 & 100\_HPS\_0   \\
28.12996 & -16.754614 & 0.96 & 13 & 10 & 0.8 & 9 & 100\_HPS\_0   \\
28.02337 & -16.698863 & 0.96 & 6 & 18 & 0.8 & 6 & 100\_HPS\_0   \\
28.44967 & -16.368691 & 0.94 & 8 & 10 & 0.6 & 7 & 100\_HPS\_0   \\
28.35613 & -16.780046 & 0.94 & 8 & 20 & 0.8 & 6 & 30\_HPS\_1 70\_HPS\_1  \\
28.14254 & -16.756250 & 0.93 & 5 & 12 & 0.4 & 6 & 100\_HPS\_0   \\
28.05023 & -16.678293 & 0.92 & 16 & 12 & 0.7 & 8 & 10\_MH\_5 90\_HPS\_0  \\
28.53401 & -16.361629 & 0.91 & 9 & 8 & 0.8 & 7 & 99\_HPS\_0 1\_PCA\_0  \\
28.48467 & -16.225677 & 0.91 & 6 & 100 & 0.1 & 7 & 30\_HPS\_0 70\_L40\_0  \\
28.20192 & -16.827965 & 0.9 & 15 & 25 & 0.8 & 9 & 100\_HPS\_0   \\
28.49411 & -16.372853 & 0.89 & 8 & 25 & 0.5 & 7 & 100\_HPS\_0   \\
28.07346 & -16.671971 & 0.89 & 8 & 10 & 0.8 & 7 & 100\_HPS\_0   \\
28.08931 & -16.658905 & 0.87 & 9 & 12 & 0.75 & 5 & 100\_HPS\_0   \\
28.37785 & -16.552871 & 0.85 & 9 & 10 & 0.9 & 5 & 100\_HPS\_0   \\
28.09282 & -16.631381 & 0.84 & 5 & 8 & 0.5 & 5 & 90\_HPS\_1.5 10\_HPS\_0  \\
28.46777 & -16.447119 & 0.84 & 8 & 25 & 0.5 & 9 & 100\_HPS\_0   \\
28.29346 & -16.377225 & 0.82 & 16 & 30 & 0.7 & 8 & 100\_HPS\_0   \\
28.55389 & -16.344736 & 0.82 & 16 & 20 & 0.6 & 6 & 95\_HPS\_0 3\_PCA\_0 2\_L27\_0 \\
28.37744 & -16.639092 & 0.82 & 4 & 18 & 0.6 & 6 & 100\_HPS\_0   \\
28.03476 & -16.637584 & 0.82 & 9 & 12 & 0.9 & 6 & 100\_HPS\_0   \\
28.47452 & -16.436175 & 0.82 & 7 & 8 & 0.5 & 7 & 100\_HPS\_0   \\
28.02908 & -16.605236 & 0.82 & 16 & 50 & 0.4 & 4 & 70\_HPS\_38 30\_HPS\_0  \\
28.12925 & -16.529170 & 0.82 & 8 & 9 & 0.5 & 5 & 100\_HPS\_0   \\
28.15614 & -16.635884 & 0.81 & 7 & 15 & 0.9 & 7 & 5\_HPS\_38 95\_HPS\_0  \\
28.50974 & -16.193821 & 0.81 & 12 & 12 & 0.7 & 6 & 100\_HPS\_0   \\
28.38366 & -16.611910 & 0.8 & 12 & 25 & 0.8 & 7 & 50\_HPS\_3 50\_HPS\_0  \\
28.35657 & -16.734756 & 0.8 & 7 & 10 & 0.7 & 7 & 100\_HPS\_3   \\
28.09959 & -16.680719 & 0.8 & 8 & 9 & 0.9 & 6 & 10\_HPS\_0 90\_HPS\_38  \\
28.38511 & -16.584514 & 0.8 & 24 & 10 & 0.9 & 9 & 100\_PCA\_0   \\
28.37792 & -16.570143 & 0.8 & 7 & 18 & 0.6 & 6 & 100\_PCA\_0   \\
28.45926 & -16.420817 & 0.78 & 8 & 12 & 0.6 & 9 & 100\_HPS\_0   \\
28.35389 & -16.373236 & 0.78 & 16 & 10 & 0.8 & 5 & 95\_HPS\_0 5\_MH\_0  \\
28.09810 & -16.617865 & 0.78 & 6 & 6 & 0.7 & 5 & 90\_HPS\_38 10\_HPS\_0  \\
28.37363 & -16.652382 & 0.78 & 6 & 10 & 0.9 & 5 & 40\_HPS\_0 60\_HPS\_2  \\
28.18275 & -16.480165 & 0.77 & 8 & 12 & 0.7 & 6 & 100\_HPS\_0   \\
28.38805 & -16.506205 & 0.75 & 10 & 9 & 0.9 & 6 & 100\_HPS\_1   \\
28.25800 & -16.426538 & 0.74 & 8 & 100 & 0.05 & 6 & 80\_HPS\_3 20\_HPS\_20  \\
28.41205 & -16.504643 & 0.74 & 10 & 28 & 0.5 & 5 & 60\_HPS\_0 40\_PCA\_0  \\
28.36623 & -16.499300 & 0.74 & 4 & 15 & 0.5 & 8 & 100\_HPS\_0   \\
28.40159 & -16.509613 & 0.73 & 4 & 18 & 0.9 & 5 & 100\_HPS\_0   \\
28.38382 & -16.596720 & 0.73 & 8 & 9 & 0.5 & 4 & 100\_HPS\_0   \\
28.15916 & -16.766020 & 0.73 & 10 & 7 & 0.6 & 6 & 97\_HPS\_0 3\_HPS\_1  \\
28.08029 & -16.680308 & 0.7 & 14 & 10 & 0.7 & 6 & 100\_HPS\_0   \\
28.33809 & -16.419592 & 0.7 & 7 & 7 & 0.9 & 5 & 95\_HPS\_0 5\_HPS\_38  \\
28.04859 & -16.660106 & 0.69 & 8 & 16 & 0.7 & 6 & 100\_HPS\_0   \\
28.33082 & -16.399790 & 0.69 & 6 & 8 & 0.9 & 5 & 100\_HPS\_0   \\
28.26909 & -16.819540 & 0.69 & 10 & 18 & 0.5 & 7 & 97\_PCA\_0 3\_L27\_0  \\
28.10252 & -16.587477 & 0.65 & 8 & 10 & 0.7 & 6 & 100\_HPS\_0   \\
28.01262 & -16.668922 & 0.65 & 12 & 8 & 0.9 & 8 & 75\_HPS\_0 25\_HPS\_3  \\
28.07459 & -16.694939 & 0.64 & 6 & 12 & 0.6 & 4 & 90\_HPS\_1.5 10\_HPS\_0  \\
28.15323 & -16.728272 & 0.63 & 6 & 12 & 0.75 & 6 & 100\_HPS\_0   \\
28.28033 & -16.409189 & 0.62 & 8 & 9 & 0.5 & 6 & 100\_HPS\_0   \\
28.11100 & -16.595715 & 0.62 & 10 & 20 & 0.7 & 7 & 40\_HPS\_38 10\_HPS\_1.5 50\_HPS\_3 \\
28.37534 & -16.584072 & 0.62 & 9 & 7 & 0.6 & 8 & 100\_PCA\_0   \\
28.37693 & -16.512386 & 0.61 & 8 & 13 & 0.7 & 8 & 100\_HPS\_0   \\
28.34693 & -16.722251 & 0.61 & 6 & 20 & 0.7 & 6 & 100\_HPS\_1   \\
28.32740 & -16.804861 & 0.61 & 8 & 4 & 0.5 & 7 & 97\_PCA\_0 3\_L27\_0  \\
28.23876 & -16.796690 & 0.6 & 8 & 14 & 0.75 & 8 & 100\_HPS\_3   \\
28.39785 & -16.554391 & 0.6 & 8 & 15 & 0.7 & 7 & 100\_HPS\_3   \\
28.44270 & -16.282202 & 0.6 & 7 & 30 & 0.95 & 7 & 90\_HPS\_0 10\_MH\_5  \\
28.39578 & -16.544884 & 0.59 & 10 & 65 & 0.6 & 9 & 80\_HPS\_0 20\_MH\_15  \\
28.29615 & -16.815608 & 0.59 & 8 & 27 & 0.4 & 7 & 97\_PCA\_0 3\_L27\_0  \\
28.43321 & -16.321034 & 0.59 & 8 & 17 & 0.8 & 6 & 100\_HPS\_0   \\
28.18253 & -16.818130 & 0.58 & 14 & 10 & 0.8 & 6 & 95\_HPS\_0 5\_L27\_0  \\
28.51595 & -16.360996 & 0.58 & 8 & 40 & 0.1 & 4 & 100\_HPS\_0   \\
28.39829 & -16.357672 & 0.57 & 4 & 10 & 0.3 & 6 & 100\_HPS\_1   \\
28.42350 & -16.318772 & 0.57 & 12 & 17 & 0.6 & 7 & 95\_HPS\_2 2.5\_MH\_2 2.5\_FLU\_2 \\
28.16574 & -16.430795 & 0.57 & 6 & 22 & 0.7 & 6 & 50\_HPS\_0 50\_HPS\_50  \\
28.04286 & -16.614751 & 0.56 & 6 & 25 & 0.8 & 6 & 80\_MH\_50 20\_HPS\_0  \\
28.42400 & -16.299263 & 0.56 & 15 & 15 & 0.8 & 7 & 80\_HPS\_0 20\_L40\_10  \\
28.05432 & -16.525011 & 0.55 & 8 & 25 & 0.8 & 4 & 100\_HPS\_0   \\
28.40166 & -16.321832 & 0.54 & 5 & 7 & 0.7 & 7 & 95\_HPS\_0 5\_MH\_5  \\
28.41454 & -16.317553 & 0.54 & 6 & 17 & 0.5 & 10 & 100\_HPS\_3   \\
28.37970 & -16.360827 & 0.53 & 30 & 20 & 0.6 & 10 & 95\_HPS\_0 5\_MH\_0  \\
28.23671 & -16.440496 & 0.53 & 16 & 20 & 0.6 & 6 & 100\_HPS\_0   \\
28.43874 & -16.369566 & 0.53 & 7 & 25 & 0.4 & 9 & 100\_HPS\_0   \\
28.42918 & -16.307490 & 0.52 & 7 & 13 & 0.9 & 7 & 95\_HPS\_0 5\_MH\_0  \\
28.49136 & -16.214020 & 0.52 & 10 & 30 & 0.4 & 7 & 80\_HPS\_0 20\_MH\_5  \\
28.06746 & -16.625952 & 0.52 & 5 & 22 & 0.8 & 8 & 40\_HPS\_38 10\_HPS\_1.5 50\_HPS\_3 \\
28.18666 & -16.765573 & 0.51 & 12 & 9 & 0.5 & 8 & 30\_HPS\_3 70\_HPS\_0  \\
28.33915 & -16.790778 & 0.51 & 8 & 7 & 0.5 & 6 & 50\_HPS\_1 50\_HPS\_38  \\
28.38759 & -16.561770 & 0.51 & 9 & 15 & 0.7 & 8 & 100\_PCA\_0   \\
28.32649 & -16.787447 & 0.51 & 8 & 12 & 0.5 & 6 & 50\_HPS\_38 50\_HPS\_0  \\
28.14390 & -16.443432 & 0.5 & 6 & 8 & 0.9 & 5 & 100\_HPS\_38   \\
28.48085 & -16.245588 & 0.5 & 30 & 30 & 0.6 & 7 & 90\_HPS\_0 10\_MH\_3  \\
28.23301 & -16.456764 & 0.5 & 8 & 10 & 0.4 & 6 & 100\_HPS\_0   \\
28.06971 & -16.510903 & 0.5 & 7 & 100 & 0.05 & 9 & 100\_L40\_100   \\
28.12227 & -16.462628 & 0.5 & 9 & 6 & 0.9 & 6 & 100\_HPS\_1   \\
28.33926 & -16.756488 & 0.48 & 8 & 15 & 0.1 & 7 & 100\_HPS\_0   \\
28.22711 & -16.782947 & 0.48 & 4 & 8 & 0.8 & 7 & 95\_HPS\_1 5\_HPS\_0  \\
28.40917 & -16.333454 & 0.48 & 5 & 10 & 0.7 & 10 & 100\_HPS\_0   \\
28.49304 & -16.354686 & 0.47 & 6 & 12 & 0.6 & 8 & 95\_HPS\_0 5\_PCA\_0  \\
28.39138 & -16.670902 & 0.47 & 6 & 15 & 0.8 & 6 & 100\_HPS\_0   \\
28.56031 & -16.218691 & 0.47 & 4 & 7 & 0.8 & 6 & 100\_HPS\_0   \\
28.40934 & -16.311675 & 0.46 & 10 & 15 & 0.5 & 9 & 100\_HPS\_0   \\
28.26815 & -16.806913 & 0.46 & 9 & 7 & 0.6 & 7 & 97\_PCA\_0 3\_L27\_0  \\
28.39667 & -16.347677 & 0.46 & 4 & 8 & 0.9 & 4 & 75\_HPS\_1 25\_HPS\_0  \\
28.34160 & -16.731272 & 0.44 & 4 & 6 & 0.7 & 7 & 80\_HPS\_1 20\_HPS\_0  \\
28.10510 & -16.559746 & 0.43 & 4 & 25 & 0.4 & 6 & 99\_HPS\_0 1\_HPS\_38  \\
28.52626 & -16.154935 & 0.43 & 10 & 11 & 0.5 & 7 & 100\_HPS\_0   \\
28.18068 & -16.792192 & 0.41 & 12 & 12 & 0.8 & 9 & 100\_HPS\_0   \\
28.39212 & -16.657583 & 0.41 & 6 & 20 & 0.6 & 4 & 100\_HPS\_0   \\
28.09897 & -16.481605 & 0.41 & 6 & 6 & 0.9 & 6 & 100\_HPS\_1   \\
28.16870 & -16.733814 & 0.4 & 5 & 30 & 0.2 & 6 & 99\_HPS\_0 1\_FLU\_38  \\
28.39388 & -16.591268 & 0.4 & 9 & 15 & 0.9 & 5 & 90\_PCA\_0 10\_HPS\_0  \\
28.14233 & -16.522501 & 0.4 & 8 & 10 & 0.5 & 5 & 100\_HPS\_0   \\
28.41704 & -16.304883 & 0.4 & 5 & 10 & 0.8 & 4 & 100\_HPS\_0   \\
28.19417 & -16.424306 & 0.4 & 9 & 40 & 0.3 & 7 & 100\_HPS\_0   \\
28.27356 & -16.384774 & 0.4 & 8 & 16 & 0.75 & 6 & 100\_HPS\_0   \\
28.38185 & -16.373466 & 0.4 & 10 & 7 & 0.7 & 9 & 90\_HPS\_2 10\_HPS\_0  \\
28.38211 & -16.534929 & 0.4 & 8 & 22 & 0.8 & 4 & 100\_HPS\_0   \\
28.36673 & -16.528122 & 0.38 & 8 & 12 & 0.6 & 5 & 100\_PCA\_0   \\
28.38796 & -16.553718 & 0.37 & 9 & 10 & 0.8 & 5 & 100\_HPS\_0   \\
28.14599 & -16.792426 & 0.37 & 48 & 40 & 0.5 & 7 & 100\_HPS\_2   \\
28.39359 & -16.649390 & 0.37 & 8 & 13 & 0.7 & 5 & 100\_HPS\_0   \\
28.40723 & -16.325813 & 0.36 & 8 & 20 & 0.7 & 4 & 100\_HPS\_0   \\
28.28347 & -16.801950 & 0.36 & 4 & 8 & 0.7 & 3 & 97\_PCA\_0 3\_L27\_0  \\
28.48806 & -16.238554 & 0.35 & 12 & 6 & 0.8 & 6 & 100\_HPS\_0   \\
28.15812 & -16.746611 & 0.35 & 7 & 20 & 0.2 & 7 & 99\_HPS\_0 1\_HPS\_1  \\
28.17986 & -16.802851 & 0.34 & 12 & 9 & 0.75 & 6 & 100\_HPS\_0   \\
28.40318 & -16.332080 & 0.33 & 8 & 10 & 0.8 & 6 & 95\_HPS\_15 5\_MH\_15  \\
28.43660 & -16.288420 & 0.32 & 7 & 60 & 0.2 & 7 & 80\_HPS\_0 20\_MH\_5  \\
28.22189 & -16.413374 & 0.31 & 8 & 12 & 0.6 & 7 & 100\_HPS\_0   \\
28.29604 & -16.565715 & 0.29 & 8 & 45 & 0.2 & 4 & 100\_FLU\_15   \\
28.57158 & -16.197841 & 0.29 & 12 & 4 & 0.95 & 6 & 100\_HPS\_0   \\
28.28710 & -16.813983 & 0.29 & 6 & 9 & 0.5 & 7 & 97\_PCA\_0 3\_L27\_0  \\
28.45347 & -16.342539 & 0.29 & 6 & 17 & 0.95 & 7 & 100\_MH\_5   \\
28.39628 & -16.641086 & 0.28 & 6 & 8 & 0.9 & 6 & 95\_HPS\_0 5\_PCA\_0  \\
28.24050 & -16.403156 & 0.25 & 12 & 10 & 0.9 & 7 & 100\_HPS\_0   \\
28.10965 & -16.470852 & 0.25 & 7 & 13 & 0.7 & 7 & 100\_HPS\_3   \\
28.14601 & -16.547024 & 0.24 & 4 & 10 & 0.6 & 6 & 95\_HPS\_0 5\_HPS\_3  \\
28.26879 & -16.426897 & 0.24 & 6 & 9 & 0.7 & 8 & 100\_HPS\_0   \\
28.10215 & -16.476654 & 0.23 & 8 & 10 & 0.9 & 7 & 100\_HPS\_2   \\
28.10540 & -16.473615 & 0.21 & 10 & 9 & 0.9 & 6 & 100\_HPS\_1   \\
28.24013 & -16.411299 & 0.21 & 5 & 5 & 0.85 & 6 & 100\_MH\_3   \\
28.24899 & -16.398963 & 0.19 & 12 & 4 & 0.7 & 7 & 100\_HPS\_0   \\
28.26028 & -16.392639 & 0.13 & 8 & 10 & 0.6 & 5 & 100\_HPS\_0   \\
28.30583 & -16.564861 & 0.06 & 6 & 100 & 0.4 & 6 & 100\_MV3\_15  
\label{inventory}
\end{longtable}

\bsp
\label{lastpage}
\end{landscape}


\end{document}